\documentclass[12pt]{article}
\usepackage{amsmath,amssymb,amsfonts,dsfont,amsthm}
\usepackage[utf8]{inputenc}
\usepackage[T1]{fontenc}
\usepackage{graphicx}
\usepackage{psfrag,epsf}
\usepackage{pifont}
\newcommand{\Cor}[1]{\textcolor{black}{#1}}
\usepackage[longnamesfirst]{natbib}
\usepackage{url} 
\usepackage{blkarray}
\usepackage{tikz}
\usepackage{placeins}
\usepackage{xcolor}
\usepackage{multirow}
\usepackage{arydshln}
\newcommand{\Corr}[1]{\textcolor{black}{#1}}

\allowdisplaybreaks
 \newtheorem{defi}{Definition}[section]
\newtheorem{theorem}[defi]{Theorem}
\newtheorem{proposition}[defi]{Proposition}
\newtheorem{conjecture}[defi]{Conjecture}
\newtheorem{corollary}[defi]{Corollary}
\newtheorem{lemma}[defi]{Lemma}
\newenvironment{demo}{\noindent{\bf Proof.}}{\qed}
\newtheorem*{remark}{Remark}
\newcommand{\blind}{1}

\newcommand{\bz}{\textbf{z}}
\newcommand{\bw}{\textbf{w}}
\newcommand{\Id}{\textbf{Id}}

\newcommand{\xmark}{\ding{55}}

\begin{document}

\def\spacingset#1{\renewcommand{\baselinestretch}%
{#1}\small\normalsize} \spacingset{1}


 \if1\blind
{
 \title{\bf Comparing high dimensional  partitions with the Coclustering Adjusted Rand Index.}
 \author{ Valerie Robert \\
 Laboratoire d'Informatique et de Math\'ematiques, Universit\'e de la R\'eunion (France),\\
valerie.robert.math@gmail.com,\\
 Yann Vasseur \\
Institut de Math\'ematiques d'Orsay, Universit\'e Paris Saclay (France),\\ 
 Vincent Brault \\
           Universit\'e Grenoble Alpes, CNRS, Grenoble INP, LJK, Grenoble (France),\\ 
           vincent.brault@univ-grenoble-alpes.fr
 \thanks{
 The authors are grateful to Gilles Celeux and Christine Keribin for initiating
this work, to Gerard Govaert for useful discussions and valuable suggestions.}
 }
 
  \date{}
 \maketitle
\hspace{-1cm}
} \fi

\if0\blind
{
 \begin{center}
 {\LARGE\bf Comparing high-dimensional partitions with the Co-clustering Adjusted Rand Index.}
\end{center}
} \fi
\hspace{-2cm}
\begin{abstract}
\Cor{We consider the simultaneous clustering of rows and columns of a matrix and more particularly the ability to measure the agreement between two co-clustering partitions.}

\Cor{The new criterion we developed is based on the Adjusted Rand Index and is called the Co-clustering Adjusted Rand Index named CARI.  We also suggest new improvements to existing criteria such as the Classification Error which counts the proportion of misclassified cells and the Extended Normalized Mutual Information criterion which is a generalization of the criterion based on mutual information in the case of classic classifications. }

\Cor{We study these criteria with regard to some desired properties deriving from the co-clustering context. Experiments on simulated and real observed data are proposed to compare the behavior of these criteria.}
\end{abstract}

\noindent%
{\it Keywords:}  Co-clustering, Adjusted Rand Index, Mutual information, Agreement, Partition 
\vfill

\newpage
\spacingset{1.45} 
\section{Introduction}
With the advent of sparse and high-dimensional
datasets in statistics, co-clustering  has become a topic of interest in recent years in many fields of applications. For example, in the context of text mining, \cite{dhillon2003information} aim at finding similar documents and their interplay with word clusters. In genomics, the objective of \cite{jagalur2007analyzing} is to identify groups of genes whose expression is linked to anatomical sub-structures in the mouse brain.  As for \cite{shan2008bayesian} or \cite{wyse2016block}, they \Corr{used co-clustering to} obtain groups of users sharing the same movie tastes and  to improve recommendation systems. Finally, we can cite a new approach in  pharmacovigilance proposed by \cite{christine2017latent} using co-clustering to simultaneously provide  clusters of individuals sharing the same drug profile along with the link between drugs and non-detected adverse events. 

Originally developed by \cite{hartigan1975Clustering}, a co-cluster analysis aims at reducing the data matrix to a simpler one \Corr{while preserving the information contained in the initial matrix} \citep{govaert2013co}. Co-clustering methods may provide a simultaneous partition of two sets $A$ (rows, observations, individuals) and $B$ (columns, variables, attributes). The major advantages of this method consist in drastically decreasing the dimension of clustering and the ease to deal with sparse data \citep{dhillon2003information}.Thanks to co-clustering methods, information about the dependence between rows and columns can be extracted. Furthermore, co-clustering results are easy to interpret through data visualization in terms of blocks.

To assess the performances of a co-clustering method, partitions obtained with the procedure need  to be evaluated. Objective criteria are therefore required to measure how close these partitions are to a reference structure. 

On the one hand, \cite{charrad2010determination} suggest a first solution and linearly extend from clus\-tering  several internal validity indices  (Dunn index, Baker and
Hubert index,  Davies and Bouldin index, Calinsky and Harabsz index, 
Silhouette criterion,  Hubert and Levin index,  Krzanowski and Lai index and the differential method). \cite{wyse2016block} also extend in the same way, through the use of an external validity index, that is based on the normalized mutual information measure discussed in \cite{vinh2010information}.

 However, simply defining a linear combination between the index for row partitions and the index for column partitions fails to preserve the co-clustering structure. \Corr{From a co-clustering} point of view, the statistical unity is a cell within a matrix which is characterized simultaneously by its row number and its column number. If the rows are considered separately from the columns through in an index, a wrongly classified cell may be counted twice.
On the other hand, \cite{Lomet2012selection} propose a distance measure particularly designed for assessing agreement between co-clustering partitions. The computation of this index is  dependent on the number of partition permutations and this property make its calculation time-consuming. More precisely, the numbers of clusters  can  barely exceed nine in each direction. Moreover, no mathematical convention is given when the number of clusters of the compared partitions is different.  

The aim of the present paper is to adapt the very popular and commonly used \textit{Adjusted Rand Index} (ARI) developed by \cite{hubert1985comparing} to a co-clustering point of view. In order to challenge other indices and tackle the problem of pairs of partitions with possibly different and high numbers of clusters, this new index takes into account the co-clustering structure while keeping its computation time efficient. 

The paper is organized as follows. In the next section, the statistical framework is introduced with \Cor{the desired properties to evaluate two co-clustering partitions. The new index  we developed, which is named the  \textit{Co-clustering Adjusted Rand Index} (CARI) is presented. Furthermore, two indices that have also been proposed for comparing co-clustering partitions are considered and examined from a co-clustering point of view; improvements are suggested each time. In Section 3, theoretical properties of the five criteria are studied. Section 4 is devoted to evaluating and comparing the criteria regarding the desired properties. In Section 5, a real observed data application is presented in which the application of CARI is illustrated.  Finally a discussion section ends this paper.}


\section{Statistical framework }

In this section, we present the co-clustering framework and the new external validity index we developed to assess co-clustering results. Other competitor criteria are also detailed and compared with the proposed criterion.

\subsection{Co-clustering context}\label{sec:co-clust}

Co-clustering is a method of cross-classification of arrays whose objective is to form a grid of blocks that are as homogeneous as possible inside and the most heterogeneous when they are taken in pairs (see Figure~\ref{Fig:LBM:Exemple}). This grid is the Cartesian product of a partition of the rows and a partition of the columns of the matrix.

\begin{figure}[!h]
\centering
\begin{tabular}{cc}
\begin{minipage}[c]{0.3\linewidth}
\begin{center}
\begin{tikzpicture}
\draw (0.5,0) grid[step=0.5] (4,5);
\draw (0,4.75) node[right]{A};
\draw (0,4.25) node[right]{B};
\draw (0,3.75) node[right]{C};
\draw (0,3.25) node[right]{D};
\draw (0,2.75) node[right]{E};
\draw (0,2.25) node[right]{F};
\draw (0,1.75) node[right]{G};
\draw (0,1.25) node[right]{H};
\draw (0,0.75) node[right]{I};
\draw (0,0.25) node[right]{J};
\draw (0.75,5) node[above]{1};
\draw (1.25,5) node[above]{2};
\draw (1.75,5) node[above]{3};
\draw (2.25,5) node[above]{4};
\draw (2.75,5) node[above]{5};
\draw (3.25,5) node[above]{6};
\draw (3.75,5) node[above]{7};
\draw[fill=black] (0.5,4.5) rectangle (1.5,5);
\draw[fill=black] (2,4.5) rectangle (2.5,5);
\draw[fill=black] (3,4.5) rectangle (3.5,5);
\draw[fill=black] (1.5,4) rectangle (2,4.5);
\draw[fill=black] (2.5,4) rectangle (3,4.5);
\draw[fill=black] (3.5,4) rectangle (4,4.5);
\draw[fill=black] (0.5,3.5) rectangle (1.5,4);
\draw[fill=black] (2,3.5) rectangle (2.5,4);
\draw[fill=black] (3,2.5) rectangle (3.5,4);
\draw[fill=black] (1,3) rectangle (1.5,3.5);
\draw[fill=black] (1.5,2) rectangle (2,2.5);
\draw[fill=black] (2.5,2) rectangle (3,2.5);
\draw[fill=black] (3.5,2) rectangle (4,2.5);
\draw[fill=black] (1,1.5) rectangle (1.5,2);
\draw[fill=black] (3,0.5) rectangle (3.5,2);
\draw[fill=black] (0.5,1) rectangle (1,1.5);
\draw[fill=black] (1.5,1) rectangle (2.5,1.5);
\draw[fill=black] (1,0.5) rectangle (1.5,1);
\draw[fill=black] (1.5,0) rectangle (2,0.5);
\draw[fill=black] (2.5,0) rectangle (3,0.5);
\draw[fill=black] (3.5,0) rectangle (4,0.5);
\end{tikzpicture}
\end{center}
\end{minipage}&
\begin{minipage}[c]{0.3\linewidth}
\begin{center}
\begin{tikzpicture}
\draw (0.5,0) grid[step=0.5] (4,5);
\draw (0,4.75) node[right]{E};
\draw (0,4.25) node[right]{D};
\draw (0,3.75) node[right]{G};
\draw (0,3.25) node[right]{I};
\draw (0,2.75) node[right]{A};
\draw (0,2.25) node[right]{C};
\draw (0,1.75) node[right]{H};
\draw (0,1.25) node[right]{B};
\draw (0,0.75) node[right]{F};
\draw (0,0.25) node[right]{J};
\draw (0.75,5) node[above]{2};
\draw (1.25,5) node[above]{6};
\draw (1.75,5) node[above]{1};
\draw (2.25,5) node[above]{4};
\draw (2.75,5) node[above]{3};
\draw (3.25,5) node[above]{5};
\draw (3.75,5) node[above]{7};
\draw[fill=black] (0.5,2) rectangle (1.5,4.5);
\draw[fill=black] (1,1.5) rectangle (2.5,3);
\draw[fill=black] (2.5,1.5) rectangle (3,2);
\draw[fill=black] (2.5,0) rectangle (4,1.5);
\draw[fill=black] (1,4.5) rectangle (1.5,5);
\draw[ultra thick,color=blue] (0.5,1.5) -- (4,1.5);
\draw[ultra thick,color=blue] (0.5,3) -- (4,3);
\draw[ultra thick,color=red] (1.5,0) -- (1.5,5);
\draw[ultra thick,color=red] (2.5,0) -- (2.5,5);
\end{tikzpicture}
\end{center}
\end{minipage}\\
\end{tabular}
\caption{\label{Fig:LBM:Exemple}Co-clustering context: original data (on the left) and reordered data in terms of  blocks (on the right).}
\end{figure}

\Corr{Co-clustering, which aims at designing a grid of blocks, ought to} be contrasted with other types of cross-classifications. For example, a nested classification aims at designing blocks within already formed blocks whereas a bi-clustering aims at designing only the most homogeneous blocks without taking into account  possible overlap (see Figure~\ref{Fig:Comp:Exemple} for schematic examples and see \cite{hartigan1975Clustering}).

\begin{figure}[!h]
\centering
\begin{tabular}{ccc}
\begin{minipage}[c]{0.3\linewidth}
\begin{center}
\begin{tikzpicture}
\draw (0.5,0) grid[step=0.5] (4,5);
\draw (0,4.75) node[right]{E};
\draw (0,4.25) node[right]{D};
\draw (0,3.75) node[right]{G};
\draw (0,3.25) node[right]{I};
\draw (0,2.75) node[right]{A};
\draw (0,2.25) node[right]{C};
\draw (0,1.75) node[right]{H};
\draw (0,1.25) node[right]{B};
\draw (0,0.75) node[right]{F};
\draw (0,0.25) node[right]{J};
\draw (0,0.25) node[right]{J};
\draw (0.75,5) node[above]{2};
\draw (1.25,5) node[above]{6};
\draw (1.75,5) node[above]{1};
\draw (2.25,5) node[above]{4};
\draw (2.75,5) node[above]{3};
\draw (3.25,5) node[above]{5};
\draw (3.75,5) node[above]{7};
\draw[fill=black] (0.5,2) rectangle (1.5,4.5);
\draw[fill=black] (1,1.5) rectangle (2.5,3);
\draw[fill=black] (2.5,1.5) rectangle (3,2);
\draw[fill=black] (2.5,0) rectangle (4,1.5);
\draw[fill=black] (1,4.5) rectangle (1.5,5);
\draw[ultra thick,color=blue] (2.5,0) -- (2.5,5);
\draw[ultra thick,color=blue] (0.5,1.5) -- (4,1.5);
\draw[ultra thick,color=blue] (1.5,1.5) -- (1.5,5);
\draw[ultra thick,color=blue] (1.5,3) -- (2.5,3);

\end{tikzpicture}
\end{center}
\end{minipage}&
\begin{minipage}[c]{0.3\linewidth}
\begin{center}
\begin{tikzpicture}
\draw (0.5,0) grid[step=0.5] (4,5);
\draw (0,4.75) node[right]{E};
\draw (0,4.25) node[right]{D};
\draw (0,3.75) node[right]{G};
\draw (0,3.25) node[right]{I};
\draw (0,2.75) node[right]{A};
\draw (0,2.25) node[right]{C};
\draw (0,1.75) node[right]{H};
\draw (0,1.25) node[right]{B};
\draw (0,0.75) node[right]{F};
\draw (0,0.25) node[right]{J};
\draw (0,0.25) node[right]{J};
\draw (0.75,5) node[above]{2};
\draw (1.25,5) node[above]{6};
\draw (1.75,5) node[above]{1};
\draw (2.25,5) node[above]{4};
\draw (2.75,5) node[above]{3};
\draw (3.25,5) node[above]{5};
\draw (3.75,5) node[above]{7};
\draw[fill=black] (0.5,2) rectangle (1.5,4.5);
\draw[fill=black] (1,1.5) rectangle (2.5,3);
\draw[fill=black] (2.5,1.5) rectangle (3,2);
\draw[fill=black] (2.5,0) rectangle (4,1.5);
\draw[fill=black] (1,4.5) rectangle (1.5,5);
\draw[ultra thick,color=blue] (0.5,1.5) rectangle (1.5,5);
\draw[ultra thick,color=red] (2.5,0) rectangle (4,2);
\draw[ultra thick,color=purple] (0.5,1.5) rectangle (3,3);
\end{tikzpicture}
\end{center}
\end{minipage}
&
\begin{minipage}[c]{0.3\linewidth}
\begin{center}
\begin{tikzpicture}
\draw (0.5,0) grid[step=0.5] (4,5);
\draw (0,4.75) node[right]{E};
\draw (0,4.25) node[right]{D};
\draw (0,3.75) node[right]{G};
\draw (0,3.25) node[right]{I};
\draw (0,2.75) node[right]{A};
\draw (0,2.25) node[right]{C};
\draw (0,1.75) node[right]{H};
\draw (0,1.25) node[right]{B};
\draw (0,0.75) node[right]{F};
\draw (0,0.25) node[right]{J};
\draw (0.75,5) node[above]{2};
\draw (1.25,5) node[above]{6};
\draw (1.75,5) node[above]{1};
\draw (2.25,5) node[above]{4};
\draw (2.75,5) node[above]{3};
\draw (3.25,5) node[above]{5};
\draw (3.75,5) node[above]{7};
\draw[fill=black] (0.5,2) rectangle (1.5,4.5);
\draw[fill=black] (1,1.5) rectangle (2.5,3);
\draw[fill=black] (2.5,1.5) rectangle (3,2);
\draw[fill=black] (2.5,0) rectangle (4,1.5);
\draw[fill=black] (1,4.5) rectangle (1.5,5);
\draw[ultra thick,color=blue] (0.5,1.5) -- (4,1.5);
\draw[ultra thick,color=blue] (0.5,3) -- (4,3);
\draw[ultra thick,color=red] (1.5,0) -- (1.5,5);
\draw[ultra thick,color=red] (2.5,0) -- (2.5,5);
\end{tikzpicture}
\end{center}
\end{minipage}\\
\end{tabular}
\caption{\label{Fig:Comp:Exemple}Three methods of cross-classification: nested classification (on the left), bi-clustering (in the middle) and co-clustering (on the right).}
\end{figure}

As explained in~\cite{brault2015co}, a difference can be made between co-clustering and the need to separately cluster the rows and columns of a matrix. In the first case, the statistical unit is the cell of the matrix and the objective is to assign it to the relevant block. In the second case, the statistical units are, on the one hand, the lines and on the other hand, the columns: the goal is then to know if the rows (and independently of the columns) are then aggregated in the suitable clusters. In Figure~\ref{Fig:erreurcoVSerreurdouble}, the differences between two co-partitions  are schematized: the yellow (resp. red) cells  are correctly classified towards rows (resp columns) but misclassified towards columns (resp rows). The orange cells are located at the intersection of rows and columns that do not belong to the same clusters in both co-partitions. In the co-clustering point of view, each yellow, red, and orange cell is misclassified as it does not belong to the same blocks (for example, some of the yellow cells belong to block $(1,1)$ for the first classification and block $(2,1)$ for the second). In the second point of view, only the orange cells which are located at the intersection are \textit{totally} misclassified. The yellow and red cells are only considered \textit{partially} misclassified since only the row or the column of which they are at the intersection is misclassified.

\begin{figure}[ht!]
\centering
\begin{tikzpicture}

    \draw[blue, ultra thick, dotted] (0,6) -- (4,6);
    \draw[purple, ultra thick] (4,6) -- (8,6);
    \draw[blue, ultra thick, dotted] (0,5.5) -- (3,5.5);
    \draw[purple, ultra thick] (3,5.5) -- (4,5.5);
    \draw[blue, ultra thick, dotted] (4,5.5) -- (6,5.5);
    \draw[purple, ultra thick] (6,5.5) -- (8,5.5);
    
    \draw[green, ultra thick, dotted] (-1,0) -- (-1,4);
    \draw[orange, ultra thick] (-1,5) -- (-1,4);
    \draw[green, ultra thick, dotted] (-0.5,0) -- (-0.5,2);
    \draw[orange, ultra thick] (-0.5,5) -- (-0.5,2);
    
    \draw (8,6) node[right]{$\bw$};
    \draw (8,5.5) node[right]{$\bw'$};
    \draw (-1,-0.13) node[below]{$\bz$};
    \draw (-0.5,0) node[below]{$\bz'$};
    
    \draw[yellow,opacity=0.3,fill] (0,2) rectangle (3,4);
    \draw[yellow,opacity=0.3,fill] (6,2) rectangle (8,4);
    \draw[red,opacity=0.3,fill] (3,0) rectangle (6,2);
    \draw[red,opacity=0.3,fill] (3,4) rectangle (6,5);
    \draw[orange,opacity=0.6,fill] (3,4) rectangle (6,2);

\draw (0,0) grid[step=1] (8,5);
\draw[ultra thick] (0,0) rectangle (8,5);
\end{tikzpicture}
\caption{\label{Fig:erreurcoVSerreurdouble}Schematic representation of the differences between two cross-classifications $(\bz, \bw)$ and $(\bz', \bw')$ with two clusters in rows and two clusters in columns: the colors and the style of the lines on the left (resp. above) represent the row (resp. columns) clusters  for each co-partition.  The yellow (resp. red) cells  are correctly classified towards rows (resp columns) but misclassified towards columns (resp rows). The orange cells are located at the intersection of rows and columns that do not belong to the same clusters in both co-partitions. }

\end{figure}

Finally, as in any mixture model, the problem of \textit{label switching} must be taken into account: if two partitions are exactly equal but the cluster numbers are mixed, we must consider that the partitions are identical. 

In conclusion, the objective of this work is therefore to propose a suitable criterion that is linked to the rate of misclassified cells of a matrix. It should penalize with the same weight every cell that does not belong to the same block of two co-partitions, and take into account the label switching problem.
The desired properties are therefore the following:

\begin{enumerate}
	\item \Cor{The index should be proportional to the number of miss-classified cells.}
	\item \Cor{The index should be symmetrical regarding two co-partitions.}
	\item \Cor{The index should vary between 0 and 1, where 1 means a perfect match and 0 leads to all worst scenarios, including independence.}
	\item \Cor{The index should be independent from label switching: if two co-partitions are equal when the labels are swaped, the score should remain unchanged.}
	\item \Cor{The index should be computable within a reasonable time.}
\end{enumerate}

\subsection{Rand index (RI) and Adjusted Rand Index (ARI)}
The index we developed further is based on commonly used distances in clustering: the Rand Index and the Adjusted Rand Index. Before introducing this new index, we shall summarize the principles and definitions of the latter criteria.

 Let two partitions be $\boldsymbol{z}=(z_1,\ldots z_i,\ldots, z_I)$  and $\boldsymbol{z'}=(z'_1,\ldots,z'_i,\ldots, z'_{I})$  on a set $A$, with Card($A$)=$I$, $\max(\boldsymbol{z})=H$ and $\max(\boldsymbol{z'})=H'$. In this article, the set $A$ will be further the set of the indices of the rows of a matrix. For example, $\boldsymbol{z}$ denotes  an external reference and $\boldsymbol{z'}$ denotes a partition derived from a clustering result.

The \textit{Rand Index} (RI) developed by \cite{rand1971objective}, is a measure of the similarity between two partitions $\boldsymbol{z}$ and $\boldsymbol{z'} $, and is defined as follows:
\begin{eqnarray*}
\frac{a+d}{a+b+c+d}=\frac{a+d}{\binom{I}{2}},
\end{eqnarray*}
where,
\begin{itemize}
\item $a$ denotes the number of pairs of elements that are placed in the same cluster in $\boldsymbol{z}$ and in the same cluster in $\boldsymbol{z'}$,
\item  $b$ denotes the number of pairs of elements in the same cluster  in $\boldsymbol{z}$ but not in the same cluster in $\boldsymbol{z'}$,
\item $c$ denotes the number of pairs of elements in the same cluster in $\boldsymbol{z'}$ but not in the same cluster in $\boldsymbol{z}$,
\item $d$ denotes the number of pairs of elements in different clusters in both partitions. 
\end{itemize}

The values $a$ and $d$ can be interpreted as agreements, and $b$ and $c$ as disagreements.
To compute all these values, a contingency table can be introduced. Let $\boldsymbol{n^{zz'}}=(n_{h,h'}^{zz'})_{H\times H'}$ be the matrix where $n_{h,h'}^{zz'}$ denotes the number of elements of set $A$ which belong to both cluster $h$ and cluster $h'$. The row and column margins $n_{h,\centerdot}^{zz'}$ and $n_{\centerdot, h'}^{zz'}$ denote respectively the number of elements in  cluster $h$ and $h'$.
We  thus find the following correspondence, \citep{albatineh2006similarity}:

\begin{itemize}

\item $a =\sum\limits_{h}^H\sum\limits_{h'}^{H'} \binom{n^{zz'}_{h,h'}}{2}=\frac{\sum\limits_{h}^H\sum\limits_{h'}^{H'} \big(n^{zz'}_{h,h'}\big)^2 -I}{2},$
\item  $b = \sum\limits_{h}^{H} \binom{n^{zz'}_{h,\centerdot}}{2}-a=\frac{\sum\limits_{h}^{H} (n^{zz'}_{h,\centerdot})^{2}-\sum\limits_{h}^H\sum\limits_{h'}^{H'} \big(n^{zz'}_{h,h'}\big)^2}{2},$
\item $c =\sum\limits_{h'}^{H'} \binom{n^{zz'}_{\centerdot,h'}}{2}-a=\frac{\sum\limits_{h'}^{H'} (n^{zz'}_{\centerdot,h'})^{2}-\sum\limits_{h}^H\sum\limits_{h'}^{H'} \big(n^{zz'}_{h,h'}\big)^2}{2},$
\item $d=\binom{I}{2}-(a+b+c)$,
\end{itemize}

This symmetric index  lies between $0$ and $1$ and takes the value of $1$ when the two partitions perfectly agree  up to a permutation. Thus, by comparing pairs of elements, this index does not need to review all the permutations of studied partitions, making its computation efficient.

The expected value of the \textit{Rand Index} for two random partitions does not take a constant value and its values are concentrated in a small interval close to $1$ (\cite{fowlkes1983method}). The \textit{Adjusted Rand Index} (ARI) proposed by \cite{hubert1985comparing} enables us to overcome such drawbacks. This corrected version  assumes the generalized hypergeometric distribution as the model of randomness, that is to say  partitions
are chosen randomly  so that the number of elements in the clusters can be fixed.
The general form of this index which is the normalized difference between the \textit{Rand Index} and its expected value under the generalized hypergeometric distribution assumption,  is as follows: 
 \vspace{-0.1cm}
\begin{eqnarray}
\text{ARI}=\frac{\text{Index-Expected Index}}{\text{MaxIndex-Expected Index}}\label{arif}.
\end{eqnarray}
The quantity  MaxIndex  in Equation (\ref{arif}) is the maximum value of the index regardless of the marginal numbers.
This index is bounded by $1$, and takes this value when the two partitions are equal up to a permutation. It can also take negative values, which corresponds to less agreement than expected by chance. 

From Equation (\ref{arif}), \cite{warrens2008equivalence} shows the ARI can be written in this way: 
\begin{eqnarray}
\text{ARI}(\boldsymbol{z},\boldsymbol{z'})&=&\frac{\sum\limits_{h,h'} \binom{n^{zz'}_{h,h'}}{2} -\sum\limits_{h} \binom{n^{zz'}_{h,\centerdot}}{2}\sum\limits_{h'}\binom{n^{zz'}_{\centerdot, h'}}{2} / \binom{I}{2}}{\frac{1}{2}\bigg[\sum\limits_{h} \binom{n^{zz'}_{h,\centerdot}}{2}+
\sum\limits_{h'}\binom{n^{zz'}_{\centerdot, h'}}{2}\bigg]-\bigg[\sum\limits_{h}\binom{n^{zz'}_{h,\centerdot}}{2}\sum\limits_{h'}\binom{n^{zz'}_{\centerdot, h'}}{2}\bigg]/\binom{I}{2}}\label{ariii}\\
&=&\frac{2(da-bc)}{d(b+c)-2 bc}\nonumber.
\end{eqnarray}
Like the RI, the ARI is symmetric, that is to say ARI$(\boldsymbol{z},\boldsymbol{z'})$=ARI$(\boldsymbol{z'},\boldsymbol{z})$. Indeed, the contingency table associated with ARI$(\boldsymbol{z'},\boldsymbol{z})$, is $t(\boldsymbol{n^{zz'}})$, where $t$ denotes the transpose of a matrix. Besides, in Equation (\ref{ariii}) in which the ARI is defined,  the margins of the contingency table act in a symmetric way. That is why, while considering $\boldsymbol{n^{zz'}}$ or its transpose matrix $t(\boldsymbol{n^{zz'}})$, the ARI remains unchanged. This remark will be particularly interesting in the next section, when  the new index we developed is studied.

\subsection{Co-clustering Adjusted Rand Index (CARI)}\label{Sec:CARI}

We extend the \textit{Adjusted Rand Index} to a co-clustering point of view in order to compare two co-clustering partitions which define blocks, and no longer clusters.

Let $\boldsymbol{z}=(z_1,\ldots,z_i,\ldots z_I)$ and $\boldsymbol{z'}=(z'_1,\ldots,z'_{i},\ldots, z'_{I})$ 
be two partitions on a set $A$. Let  $\boldsymbol{w}=(w_1,\ldots,w_{j},\ldots, w_J)$ and $\boldsymbol{w'}=(w'_1,\ldots,w'_{j},\ldots, w'_{J})$ be two partitions 
on a set $B$. $(\boldsymbol{z},\boldsymbol{w})$  and $(\boldsymbol{z'},\boldsymbol{w'})$ are two co-clustering partitions on the set  $A\times B$ where an observation is denoted by $x_{ij},i=1,\ldots,I;j=1,\ldots,J$, with Card($A\times B)=I\times J$. Notice that $\boldsymbol{z}$ and $\boldsymbol{z'}$ are  called row partitions. Similarly, $\boldsymbol{w}$ and $\boldsymbol{w'}$ are called column partitions. We also define the associated binary matrices $z=(z_{ih})_{I\times H}$ and $w=(w_{j\ell})_{J\times L}$ which $z_{ih}$ indicates if the row $i$ belongs to cluster $h$ and  $w_{jl}$ indicates if the column $j$ belongs to cluster $\ell$. We use the same notations to denote the vector or the matrix, as it is commonly used in mixture models.
 \begin{defi}\label{def1}

  The contingency table $\boldsymbol{n^{zwz'w'}}=(n_{p,q}^{zwz'w'})$ where $1\leq  p \leq HL$ and $ 1\leq q\leq H'L'$, is defined so that $n_{p,q}^{zwz'w'}$ denotes the number of observations of  set $A\times B$ 
belonging to  block $p$ (related to a pair $(h,\ell)$ defined by $(\boldsymbol{z},\boldsymbol{w})$)  and  block  $q$ (related to pair $(h',\ell')$ defined by $(\boldsymbol{z'},\boldsymbol{w'})$). 
\end{defi}

The contingency table can be seen as a block matrix which consists of $H\times H'$ blocks of size $L\times L'$ (see Table \ref{aricroisee}).
\begin{center}
  \begin{table}[ht!]
 \[\begin{pmatrix}
  n^{zwz'w'}_{1,1}&n^{zwz'w'}_{1,2}&\ldots  &n^{zwz'w'}_{1,L'}&\ldots&n^{zwz'w'}_{1,(H'\text{-}1) L'+1}&\ldots & \ldots&n^{zwz'w'}_{1, H'L'}\\
 n^{zwz'w'}_{2,1}&n^{zwz'w'}_{2,2}&\ldots  &n^{zwz'w'}_{2,L'}&\ldots&n^{zwz'w'}_{2,(H'\text{-}1) L'+1}&\ldots & \ldots&n^{zwz'w'}_{2, H' L'}\\
  \vdots&\vdots&\ddots&\vdots&&\vdots&\vdots&\ddots&\vdots\\
   n^{zwz'w'}_{L,1}&n^{zwz'w'}_{L,2}&\ldots  &n^{zwz'w'}_{L,L'}&\ldots& n^{zwz'w'}_{L,(H'\text{-}1) L'+1}&\ldots & \ldots&n^{zwz'w'}_{L, H' L'} \\
  \vdots&\vdots&\ddots&\vdots&&\vdots&\vdots&&\vdots\\
 \vdots&\vdots&\ddots&\vdots&&\vdots&\vdots&&\vdots\\
   n^{zwz'w'}_{(H\text{-}1) L+1,1}&n^{zwz'w'}_{(H\text{-}1) L+1,2}&\ldots  &n^{zwz'w'}_{(H\text{-}1) L+1,L'}&\ldots& n^{zwz'w'}_{(H\text{-}1) L+1,(H'-1) L'+1}&\ldots & \ldots&n^{zwz'w'}_{(H\text{-}1) L+1, H' L'} \\
  \vdots&\vdots&\ddots&\vdots&&\vdots&\vdots&\ddots&\vdots\\
     n^{zwz'w'}_{HL,1}&n^{zwz'w'}_{HL,2}&\ldots  &n^{zwz'w'}_{HL,L'}&\ldots& n^{zwz'w'}_{HL,(H'\text{-}1) L'+1}&\ldots & \ldots&n^{zwz'w'}_{HL, H' L'} 
 \end{pmatrix}\]
\caption{Contingency table to compare two pairs of co-clustering partitions.}\label{aricroisee}
\end{table}
\end{center}
Notice that a bijection can be defined between the index $p$ of the rows of the contingency table, and the block $(h,\ell)$ defined by $(\boldsymbol{z},\boldsymbol{w})$ (see Appendix A.1).

An analogous correspondence is defined for the index $q$ and the block $(h',\ell')$ defined by $(\boldsymbol{z'},\boldsymbol{w'})$. Thus the notation ($h_p$ $\ell_p$) and ($h'_q$ $\ell'_q$) could be used.
We will see afterwards, this trick  enables us to describe $\boldsymbol{n^{zwz'w'}}$ in a convenient way.

\begin{defi}
Given two co-clustering partitions $(\boldsymbol{z},\boldsymbol{w})$ and $(\boldsymbol{z'},\boldsymbol{w'})$ and their corresponding contingency table $\boldsymbol{n^{zwz'w'}}$ specified in Definition  \ref{def1}, the \textit{Co-clustering Adjusted Rand Index} (CARI) is defined as follows:

 \begin{eqnarray*}
 \text{CARI}((\boldsymbol{z,w}),(\boldsymbol{z',w'}))=\frac{\sum\limits_{p,q} \binom{n_{p,q}^{zwz'w'}}{2} -\sum\limits_{p} \binom{n^{zwz'w'}_{p,\centerdot}}{2}\sum\limits_{q}\binom{ n^{zwz'w'}_{\centerdot, q}}{2} / \binom{I\times J}{2}}{\frac{1}{2}\bigg[\sum\limits_{p} \binom{ n^{zwz'w'}_{p,\centerdot}}{2}+
\sum\limits_{q}\binom{n^{zwz'w'}_{\centerdot, q}}{2}\bigg]-\bigg[\sum\limits_{p}\binom{n^{zwz'w'}_{p,\centerdot}}{2}\sum\limits_{q}\binom{ n^{zwz'w'}_{\centerdot, q}}{2}\bigg]/\binom{I\times J}{2}}.\label{sym}
\end{eqnarray*}
\end{defi}
Like the ARI, this index is symmetric and takes the value of $1$ when the pairs of co-partitions perfectly agree up to a permutation.
But unlike the index proposed in \cite{Lomet2012selection} with which, we will compare our work in Section 3 and 4, no mathematical convention is needed when the number of clusters is different in partitions. Moreover, it does not based on the permutations of partitions and can therefore be easily computed even if the number of row clusters or column clusters exceeds nine. That said, the  complexity to compute $\boldsymbol{n^{zwz'w'}}$ is still  substantial.

Fortunately, we manage to exhibit a link between $\boldsymbol{n^{zwz'w'}}$, $\boldsymbol{n^{zz'}}$ and $\boldsymbol{n^{ww'}}$ which  makes the computation of the CARI much faster and competitive when the numbers of clusters are high. This link is presented in the next theorem. 

\begin{theorem}\label{th:kro}

We have the following relation, 
\begin{eqnarray}
\boldsymbol{n^{zwz'w'}}=  \boldsymbol{n^{zz'}}\otimes \boldsymbol{n^{ww'}}, \label{kron}
\end{eqnarray}

 where $\otimes$ denotes the Kronecker product between two matrices, $(\boldsymbol{z},\boldsymbol{w})$ and $(\boldsymbol{z'},\boldsymbol{w'})$ are two co-clustering partitions, $\boldsymbol{n^{zwz'w'}}$ is the corresponding contingency table specified in Definition  \ref{def1}, $\boldsymbol{n^{zz'}}$ and $\boldsymbol{n^{ww'}}$ are defined in Section 2.2.

\end{theorem}
The proof of this theorem is presented in Appendix A.1.

Thanks to this property, the contingency table  $\boldsymbol{n^{zwz'w'}}$ can be computed more efficiently as we will see in the next sections. Moreover, even if  the Kronecker product is not commutative, it behaves well with both the transpose operator and the margins of a matrix (sum of the observations according to the rows or the columns) and the initial properties of CARI are kept:

\begin{corollary}\label{cor:kronecker}
\begin{enumerate}
\item For all $(p,q)\in\{1,\ldots,HL\}\times \{1,\ldots,H'L'\}$, we have the relations between the margins,
\[n^{zwz'w'}_{\centerdot, q}=  n^{zz'}_{\centerdot h'_q}\otimes n^{ww'}_{\centerdot, \ell'_q}
 \text{ and } n^{zwz'w'}_{ p, \centerdot }=  n^{zz'}_{h_p, \centerdot }\otimes n^{ww'}_{\ell_p, \centerdot }\]
 \Cor{where $(h_p,\ell_p)$ (resp. $(h'_q,\ell'_q)$) are the coordinates in the co-clustering partition $(\bz,\bw)$ (resp. $(\bz',\bw')$) associated with the block $p$ (resp. $q$).}
\item The CARI criterion associated with the contingency table $\boldsymbol{n^{zwz'w'}}$ defined as in Equation (\ref{kron}) remains symmetric, that is to say,
\[\text{CARI}((\boldsymbol{z,w}),(\boldsymbol{z',w'}))=\text{CARI}((\boldsymbol{z',w'}),(\boldsymbol{z,w})).\]
\end{enumerate}
\end{corollary}

 The proof of this corollary is presented in Appendix A.2.

In the further sections, the contingency table $\boldsymbol{n^{zwz'w'}}$ is now defined by Equation (\ref{kron}).\\

To illustrate the properties of CARI, we developed several simple examples.
Consider the following pairs of co-partitions $(\boldsymbol{z,w})=((1, 1, 3, 2), (1, 2, 1, 4, 3 ))$ and $(\boldsymbol{z',w'}) = (( 2, 2, 1, 3  ),(2, 1, 2, 3, 4 ))$ which are equal up to a permutation. The contingency table (see Table \ref{ex1}) associated with CARI$((\boldsymbol{z,w}),(\boldsymbol{z',w'}))$ has a size of $(3\times 4,3\times 4)$. 
Thus, the CARI$((\boldsymbol{z,w}),(\boldsymbol{z',w'}))$ criterion behaves in the desired way and is equal  $\frac{11- 121/190}{1/2\times 22-121/190}=1$.

\begin{table}[ht!]
\small{
\begin{tabular}{c|c c c c c c c c c c c c}
Block& $(1,1)$ &$(1,2)$&$(1,3)$& $(1,4)$&$(2,1)$&$(2,2)$&$(2,3)$&$(2,4)$&$(3,1)$&$(3,2)$& $(3,3)$&$(3,4)$\\
\hline
$(1,1)$&0&	0	&0&	0&	0	&4	&0	&0&	0&	0&	0&	0\\
$(1,2)$&0	&0	&0	&0	&2	&0	&0	&0	&0	&0	&0	&0\\
$(1,3)$&0	&0&	0&	0&	0&	0&	0&	2&	0&	0&	0&	0\\
$(1,4)$&0	&0	&0	&0	&0	&0	&2	&0&	0	&0&	0	&0\\
$(2,1)$&	0&	0	&0&	0&	0&	0&	0	&0	&0	&2&	0&0\\
$(2,2)$&0&	0&	0&	0	&0	&0&	0&	0&	1&	0&	0&	0\\
$(2,3)$&0	&0&	0&	0&	0&	0&	0&	0&	0&	0&	0&	1\\
$(2,4)$&0&	0&	0&	0	&0	&0	&0	&0	&0	&0	&1	&0\\
$(3,1)$&0&	2&	0&	0&	0&	0&	0&	0&	0&	0&	0&	0\\
$(3,2)$&1&	0&	0&	0&	0&	0&	0&	0&	0&	0&	0&	0\\
$(3,3)$&0	&0	&0&	1&	0	&0&	0&	0	&0	&0	&0	&0\\
$(3,4)$&0	&0	&1&	0&	0&	0&	0&	0&	0&	0&	0&	0\\
\end{tabular}}
 \vspace{0.01cm}
\caption{Initial contingency table $\boldsymbol{n^{zwz'w'}}$ (see Definition 2.1).}\label{ex1}
\end{table}

 Let us now consider the following co-partitions $(\boldsymbol{z,w})=((1, 2, 2, 2, 1), (1, 1, 2, 1, 1, 2 ))$ and $(\boldsymbol{z',w'}) = (( 1, 1, 2, 1, 1  ),( 1, 1, 2, 1, 3, 2 ))$.  Note that partitions $\boldsymbol{w}$ and $\boldsymbol{w'}$ do not have the same number of clusters.
 The initial contingency tables related to ARI$(\boldsymbol{z,z'})$, ARI$(\boldsymbol{w,w'})$ and CARI$((\boldsymbol{z,w}),(\boldsymbol{z',w'}))$ are described in Table \ref{ex2z}  and  in Table \ref{ex2zw}. Each cell of Table \ref{ex2zw} represents the number of observations which belongs to both blocks in the first and the second co-clustering partitions, for all blocks. As announced, we observe that $ \boldsymbol{n^{zwz'w'}}=  \boldsymbol{n^{zz'}} \otimes \boldsymbol{n^{ww'}}.$ and the value of CARI is equal approximatively 0.2501.
 

 \begin{table}[ht!]
 
   \begin{minipage}[c]{.5\linewidth}
   \begin{center}
\begin{tabular}{c|c c|c}
Cluster&1&2&Margin\\
\hline
1& 2 &0&2\\
2&2&1&3\\
\hline
Margin&4&1&5\\
\end{tabular}

\end{center}
\end{minipage}\hfill
\begin{minipage}[c]{.5\linewidth}
   \begin{center}
   \begin{tabular}{c|c c c|c}
Cluster&1&2&3&Margin\\
\hline
1& 3 &0&1&4\\
2&0&2&0&2\\
\hline
Margin&3&2&1&6\\
\end{tabular}
\end{center}
\end{minipage}
 \vspace{0.02cm}
\caption{Contingency tables $\boldsymbol{n^{zz'}}$(on the left) and $\boldsymbol{n^{ww'}}$(on the right).}\label{ex2z}
 \end{table}
  
\FloatBarrier
 
 \begin{table}[ht!]
 \begin{center}
\begin{tabular}{c|c c c c c c|c}
Block&$(1,1)$&$(1,2)$&$(1,3)$&$(2,1)$&$(2,2)$&$(2,3)$&Margin\\
\hline
$(1,1)$&6	&0&	2&	0&	0&	0&8\\
$(1,2)$&0&	4	&0&	0&	0&	0&4\\
$(2,1)$&6&	0	&2	&3	&0&	1&12\\
$(2,2)$&0&	4&	0&	0&	2&	0&6\\
\hline
Margin&12&8&4&3&2&1&30\\
\end{tabular}
\end{center}
 \vspace{0.1cm}
\caption{  Initial contingency table $\boldsymbol{n^{zwz'w'}}$ (see Definition 2.1).}\label{ex2zw}
 \end{table}




\subsection{Classification error}

 The classification distance presented in \cite{Lomet2012selection} aims at studying the misclassification rate of the observations in the blocks. \Corr{Given two co-clustering partitions, $(\boldsymbol{z},\boldsymbol{w})$ and $(\boldsymbol{z'},\boldsymbol{w'})$, and w}ith the notations used in Section 2.3, the classification error is defined as follows, 
\begin{eqnarray*}
\hspace{-0.7cm}\text{dist}_{\scriptsize{(I,H)\times(J,L)}}((\boldsymbol{z},\boldsymbol{w}),(\boldsymbol{z'},\boldsymbol{w'}))&\hspace{-0.3cm}=\hspace{-0.5cm}\min\limits_{\sigma\in \mathfrak{S}(\{1,...,H\})}\min\limits_{\tau\in \mathfrak{S}(\{1,...,L\})}\left(1- \frac{1}{I\times J}\sum\limits_{i,j,h,\ell}z_{ih}z'_{i\sigma(h)}w_{j\ell}w'_{j\tau(\ell)}\right),
\end{eqnarray*}
where $\mathfrak{S}(\{1,...,H\})$ denotes the set of permutations on the set $\{1,\ldots,H\}$.

\begin{remark}
\Cor{The previous definition assumes that the two classifications have the same number of clusters; if not, just assume that there are empty clusters and take $H=H'$.}
\end{remark}

The classification error (CE) is then defined when the cost function measures the difference between \Cor{two co-clustering partitions $(\boldsymbol{z},\boldsymbol{w})$ and $(\boldsymbol{z'},\boldsymbol{w'})$}:
\begin{eqnarray*}
\text{CE}((\boldsymbol{z},\boldsymbol{w}),(\boldsymbol{z'},\boldsymbol{w'}))&=&\text{dist}_{\scriptsize{(I,H)\times(J,L)}}((\boldsymbol{z},\boldsymbol{w}),(\boldsymbol{z'},\boldsymbol{w'})).
\end{eqnarray*} 

The classification error produces a value between $0$ and $1$. Thus, the observation $x_{ij}$ is not in  block $(h,\ell)$ if row $i$ is not in  cluster $h$ \textit{or} if  column $j$ is not in cluster $\ell$. When a column is improperly classified, all the cells of this column are penalized, and the classification error is increased by $\frac{1}{J}$.

Furthermore, the  distance related to the row partitions can also be defined as follows:
\begin{equation}\label{Eq:dist_CE}
\text{\small dist}_{I,H}(\boldsymbol{z},\boldsymbol{z'})=1-\max_{\sigma\in \mathfrak{S}(\{1,...,H\})} \frac{1}{I}\sum_{i,h}z_{ih}z'_{i\sigma(h)}.
\end{equation}
 
 When the partitions do not include the same number of clusters, a suitable convention we can propose, is to consider  $H$ as the maximal number of clusters and the created additional clusters are assumed to be empty. The computation of this distance when $H$ is higher than nine, remains difficult as the order of the set $\mathfrak{S}(\{1,...,H\})$ is $H!$.

In a symmetric way, the  distance related to the column partitions is denoted by
$\text{dist}_{\scriptsize{J,L}}$.

\cite{Lomet2012selection} shows that the classification error  could be expressed  in terms of the distance related to the row partitions and the distance related to the column partitions: 
\begin{eqnarray}
\hspace{-0.4cm}\text{dist}_{\scriptsize{(I,H)\times(J,L)}}((\boldsymbol{z},\boldsymbol{w}),(\boldsymbol{z'},\boldsymbol{w'}))&=&\text{dist}_{\scriptsize{I,H}}(\boldsymbol{z},\boldsymbol{z'})+\text{dist}_{\scriptsize{J,L}}(\boldsymbol{w},\boldsymbol{w'})\nonumber\\
&-&\text{dist}_{\scriptsize{I,H}}(\boldsymbol{z},\boldsymbol{z'})\times\text{dist}_{\scriptsize{J,L}}(\boldsymbol{w},\boldsymbol{w'}).\label{lom}
\end{eqnarray} 

\Corr{As we prove in Section~\ref{sec:theory}, computing a quantity involving the set of permutations is impossible as soon as $H$ or $L$ are greater than 10. However, this problem can be seen as an \textit{assignment problem} which allows the use of optimization algorithms such as the \textit{Hungarian algorithm} proposed by~\cite{kuhn1955hungarian}. Indeed, we observe that we can rewrite Equation~\eqref{Eq:dist_CE} in the following way:
\begin{equation}\label{Eq:assignement}
\sum_{i,h}z_{ih}z'_{i\sigma(h)}=\sum_{h}\boldsymbol{n}_{h\sigma(h)}^{\boldsymbol{zz'}}.
\end{equation}
Optimizing the criterion in Equation~\eqref{Eq:dist_CE} means to search for a permutation of the columns of the matrix~$\boldsymbol{n}^{\boldsymbol{zz'}}$ so that the diagonal can have the largest possible sum. The rows in the contingency matrix~$\boldsymbol{n}^{\boldsymbol{zz'}}$ can be interpreted as agents, and the columns as tasks to be performed; each cell  represents each agent's capacity to perform a task. The aim of the assignment problem is to assign one task to an agent so that the overall capacities can be optimized.}

\Cor{Thanks to Reformulation~\eqref{Eq:assignement} of the problem, we can improve the conjecture proposed by~\cite{brault2014estimation} demonstrated in Section~\ref{sec:boundaries}: the $\text{CE}$ criterion has a strictly positive lower bound depending on $H$ and $L$. We propose to redefine the criterion in order to satisfy the desired characteristics:
\begin{equation*}
\text{NCE}((\boldsymbol{z},\boldsymbol{w}),(\boldsymbol{z'},\boldsymbol{w'}))=1-\frac{\text{dist}_{\scriptsize{(I,H)\times(J,L)}}((\boldsymbol{z},\boldsymbol{w}),(\boldsymbol{z'},\boldsymbol{w'}))}{\frac{1}{H}+\frac{1}{L}-\frac{1}{HL}}
\end{equation*}
where $\text{dist}_{\scriptsize{(I,H)\times(J,L)}}$ is defined in Equation~\eqref{lom}.
}

\begin{remark}
\Cor{Another solution could have been to normalize the distance related to the row partition and the distance related to the column partition independently, then to use Formula~\eqref{lom}. However, this choice would create an asymmetry between the roles of the row and column classifications. We rather aim at designing a criterion which counts the number of misclassified cells and is between the value of 0 and the value of 1.}
\end{remark}


\subsection{Extended Normalized Mutual Information }\label{sec:critENM}

Information theory is another approach to compare the
difference in information between two partitions. The Normalized Mutual Information (see for example \cite{linfoot1957informational}, \cite{knobbe1996analysing}, \cite{quinlan1986induction}) is a popular measure in this field. Originally, the Mutual Information (MI) between two partitions $\boldsymbol{z}=(z_1,\ldots, z_I)$  and $\boldsymbol{z'}=(z'_1,\ldots, z'_{I})$  on a same set $A=\{O_1,\ldots,O_I\}$ (with $\max(z)=H$ and $\max(z')=H'$) is defined as follows: 
\[\text{MI}(\boldsymbol{z},\boldsymbol{z'})=\sum\limits_{h,h'} P_{h,h'}\log\big(\frac{P_{h,h'}}{P_{h}P_{h'}}\big),\]

\[\text{where}, \quad P_{h,h'}=\frac{1}{I}\sum_{i}^I\mathds{1}_{\{z_i=h,z'_{i}=h'\}}, \quad P_{h}=\frac{1}{I}\sum_{i}^I\mathds{1}_{\{z_i=h\}}\quad \text{ and } \quad P_{h'}=\frac{1}{I}\sum_{i}^I\mathds{1}_{\{z'_{i}=h'\}}.\] 
As the measure has no specific upper bound, the mutual information is normalized. This task may be performed in many ways (see \cite{pfitzner2009characterization}), but  \cite{wyse2016block} choose the following normalization to define their co-clustering index:

\[\text{NMI}(\boldsymbol{z},\boldsymbol{z'})=\frac{\text{MI}(\boldsymbol{z},\boldsymbol{z'})}{\max(\mathcal{H}(\boldsymbol{z}),\mathcal{H}(\boldsymbol{z'}))},\] 
\[\text{where,\quad }\mathcal{H}(\boldsymbol{z})=-\sum_{h} P_h\log P_h,\text{ and }  \mathcal{H}(z')=-\sum_{h'} P_{h'}\log P_{h'}.\]
The proposed measure by  \cite{wyse2016block}  to compare two co-clustering partitions $(\boldsymbol{z}=(z_1,\ldots, z_I)$, $\boldsymbol{w}=(w_1,\ldots, w_J)$)  and ($\boldsymbol{z'}=(z'_1,\ldots, z'_{I})$, $\boldsymbol{w'}=(w'_1,\ldots, w'_{J})$) on a set $A\times B$ (with Card$(A)=I$ and Card$(B)=J$) is based on a linear combination of the normalized mutual information of  $\boldsymbol{z}$ and $\boldsymbol{z'}$, and the normalized mutual information of $\boldsymbol{w}$ and $\boldsymbol{w'}$:
\[\text{ENMI}((\boldsymbol{z},\boldsymbol{w}),(\boldsymbol{z'},\boldsymbol{w'}))= \text{NMI}(\boldsymbol{z},\boldsymbol{z'})+\text{NMI}(\boldsymbol{w},\boldsymbol{w'}).\]
The maximum value of this index is equal 2 when the partitions perfectly match up to a permutation and is equal 0 when the correspondence between them is extremely weak (see Corollary~\ref{cor:borne:ENMI}). However, extending the index in this way fails to preserve the co-clustering structure \Cor{(see Section~\ref{sec:CritFacing}). We therefore propose an adaptation according to the philosophy of the latent block model. 
We propose the following formula for the new criterion we developed, given two co-partitions, $(\boldsymbol{z},\boldsymbol{w})$ and $(\boldsymbol{z}',\boldsymbol{w}')$:
\begin{equation*}
\text{coMI}((\boldsymbol{z},\boldsymbol{w});(\boldsymbol{z}',\boldsymbol{w}'))=\sum_{p,q}\frac{n_{p,q}^{zwz'w'}}{IJ}\log\left(\frac{n_{p,q}^{zwz'w'}IJ}{n^{zwz'w'}_{ p, \centerdot }n^{zwz'w'}_{ \centerdot,q }}\right)
\end{equation*}
and its standardized version is as follows:
\begin{equation*}
\text{coNMI}((\boldsymbol{z},\boldsymbol{w});(\boldsymbol{z}',\boldsymbol{w}'))=\frac{\text{coMI}((\boldsymbol{z},\boldsymbol{w});(\boldsymbol{z}',\boldsymbol{w}'))}{\max(\mathcal{H}(\boldsymbol{z},\boldsymbol{w}),\mathcal{H}(\boldsymbol{z}',\boldsymbol{w}'))}
\end{equation*}
\begin{eqnarray*}
\text{where}&&\mathcal{H}(\boldsymbol{z},\boldsymbol{w})=-\sum_{p} \frac{n^{zwz'w'}_{ p, \centerdot }}{IJ}\log \left(\frac{n^{zwz'w'}_{ p, \centerdot }}{IJ}\right)\\
\text{ and}&& \mathcal{H}(\boldsymbol{z}',\boldsymbol{w}')=-\sum_{q} \frac{n^{zwz'w'}_{  \centerdot,q }}{IJ}\log \left(\frac{n^{zwz'w'}_{ \centerdot,q }}{IJ}\right).\\
\end{eqnarray*}
Thanks to Theorem~\ref{th:kro} and Corollary~\ref{cor:kronecker}, the criterion can be computable in polynomial time.
}

\begin{proposition}\label{prop:MI=coMI}
\Cor{Given two co-clustering partitions $(\boldsymbol{z},\boldsymbol{w})$ and $(\boldsymbol{z'},\boldsymbol{w'})$, we have:
\[\text{coMI}((\boldsymbol{z},\boldsymbol{w});(\boldsymbol{z}',\boldsymbol{w}'))=\text{MI}(\boldsymbol{z},\boldsymbol{z'})+\text{MI}(\boldsymbol{w},\boldsymbol{w'}).\]}
\end{proposition}

\Cor{The proof is based on Theorem~\ref{th:kro} and Corollary~\ref{cor:kronecker}}.

\begin{remark}
\Cor{The difference between criteria $ENMI/2$ and $coNMI$ is only based on the normalization; in particular, as $\mathcal{H}(\boldsymbol{z},\boldsymbol{w})=\mathcal{H}(\boldsymbol{z})+\mathcal{H}(\boldsymbol{w})$, the criteria are equal as soon as $\mathcal{H}(\boldsymbol{z})=\mathcal{H}(\boldsymbol{z'})=\mathcal{H}(\boldsymbol{w})=\mathcal{H}(\boldsymbol{w'})$.}
\end{remark}

\section{Theoretical properties of the indices}\label{sec:theory}

In this section,  we present theoretical properties of the indices such as the complexity and the boundaries of each criterion.

\subsection{Complexities}\label{sec:Complexities}
In this section, we present the complexity for the computation of each criterion.
\begin{proposition}
The complexities for the computation of each criterion are the following:
\begin{itemize}
    \item The CARI criterion has the complexity \Cor{$\mathcal{O}\left(\max\left(HH'LL',I,J\right)\right)$.}
    \item The ENMI \Cor{and coNMI criteria have} the complexity \Cor{$\mathcal{O}\left(\max\left(HH',I,LL',J\right)\right)$.}
    \item The CE criterion has the complexity $\mathcal{O}\left(\max\left(H!HI,L!LJ\right)\right)$ where $H!$ is the factorial of $H$.
    \item \Cor{The NCE criterion has the complexity $\mathcal{O}\left(\max(H^4,H'^4,I,L^4,L'^4,J)\right)$.}
\end{itemize}
\end{proposition}
The proofs are available in Appendix~\ref{section:Appendix:Complexity}. In particular, we can observe that $CE$ is usually incalculable as soon as $H$ or $L$ is greater than 10 ($10!\approx3\,628\,800$). Note that most of the steps in these procedures can be easily parallelized.

\begin{remark}
\Cor{For the complexity of the NCE criterion, the used complexity  for the Hungarian algorithm is that proposed by~\cite{kuhn1955hungarian}. In some cases, this complexity can be improved\footnote{\Cor{In the simulations, we use the \url{R}-package \url{lpSolve} proposed by~\cite{Berkelaar2020lpSolve} available on CRAN.}}.}
\end{remark}

\subsection{Boundaries}\label{sec:boundaries}
In this part, we study the boundaries of each criterion. For the CARI criterion, we only propose an upper bound of the lower bound and we conjecture that it is the upper bound too.

\begin{proposition}
If the upper bound is denoted  by $m_{CARI}$ for every partition $\bz$, $\bz'$, $\bw$ and $\bw'$, we have:
\begin{eqnarray*}
m_{CARI}&\leq&\frac{\frac{IJ}{HH'LL'}-1-\left(\frac{IJ}{HL}-1\right)\left(\frac{IJ}{H'L'}-1\right)\times\frac{1}{IJ-1}}{\frac{IJ}{2HL}+\frac{IJ}{2H'L'}-1-\left(\frac{IJ}{HL}-1\right)\left(\frac{IJ}{H'L'}-1\right)\times\frac{1}{IJ-1}}\\
\end{eqnarray*}
\end{proposition}
The proof is available in Appendix~\ref{section:Appendix:Boundarie}. From this result, we conjecture that:

\begin{conjecture}
For every partition $\bz$, $\bz'$, $\bw$ and $\bw'$, we have
\begin{eqnarray*}
\frac{HL+H'L'-HLH'L'}{IJ\left(\frac{HL+H'L'}{2}-1\right)+HL+H'L'-HLH'L'}+o(1)\leq CARI((\bz,\bz'),(\bw,\bw'))&\leq&1\\
\end{eqnarray*}
\end{conjecture}
In particular, if $H$, $H'$, $L$ and $L'$ are fixed, the lower bound is negative and tends to $0$ when $IJ$ tends to $+\infty$.

The following boundaries of the ENMI criterion are as follows:
\begin{proposition}
For every partition $\bz$ and $\bz'$, we have
\[0\leq\text{MI}(\bz,\bz')\leq\min\left(\mathcal{H}(\bz),\mathcal{H}(\bz')\right).\]
\end{proposition}
A proof is avaiblable in Appendix~\ref{section:Appendix:Boundarie}. This result can be obtained as soon as we consider that the $MI$ is a Kullback-Leibler divergence between the empirical distribution of the joint distribution and the product of the empirical marginal distributions. With this result, we can obtain the boundaries proposed by \cite{wyse2016block} in their article:
\begin{corollary}
For every partition $\bz$, $\bz'$, $\bw$ and $\bw'$, we have
\[0\leq\text{EMMI}((\bz,\bw),(\bz',\bw'))\leq2.\]
\end{corollary}

Finally, we also have boundaries of the  CE criterion:

\begin{proposition}
\Cor{ In the case where $H=\max(H,H')$ and for each partitions $\bz$ and $\bz'$, we have:
 \[0\leq\text{\small dist}_{I,H}(\boldsymbol{z},\boldsymbol{z'})\leq \frac{H-1}{H}.\]
Moreover, the upper limit is reached as soon as there are $H$ permutations $\left(\sigma_1,\ldots,\sigma_H\right)$ so that $\left(\boldsymbol{n}_{h,\sigma_{h'}\left(h\right)}^{\boldsymbol{zz'}}\right)_{1\leq h,h'\leq H}$ forms a partition of the cells of the matrix~$\boldsymbol{n}^{\boldsymbol{zz'}}$ and the values $\sum_{h=1}^H\boldsymbol{n}_{h,\sigma_{h'}\left(h\right)}^{\boldsymbol{zz'}}$ are equal.}
\end{proposition}
\Cor{We can refine the result when $ I $ is not divisible by $H^2$ (see for example Appendix~B.1 from \cite{brault2014estimation}). We propose a complete proof based on Equation~\eqref{Eq:assignement} in Appendix~\ref{section:Appendix:Boundarie}.}
\begin{corollary}\label{cor:borne:ENMI}
\Cor{For every partition $\bz$, $\bz'$, $\bw$ and $\bw'$ such that $H\geq H'$ and $L\geq L'$, we have:
\begin{eqnarray*}
1-\frac{1}{H}-\frac{1}{L}+\frac{1}{HL}\leq &1-\text{CE}((\bz,\bw),(\bz',\bw'))&\leq 1\\
\text{and }0\leq &\text{NCE}((\bz,\bw),(\bz',\bw'))&\leq 1\\
\end{eqnarray*}
where 1 is a perfect match and the lower bound is reached by all worst scenarios, including independence.}
\end{corollary}

\section{Criteria for the desired characteristics}

\Cor{In this section, we consider each point of the desired characteristics set out in Section~\ref{sec:co-clust}.}

\subsection{Criteria facing the co-clustering context}\label{sec:CritFacing}


Co-clustering a matrix and clustering its rows and its columns in a separated way are two different tasks: in co-clustering methods, the statistical unity is a cell of the matrix and no longer a row or a column. A cell is misclassified as soon as it does not belong to the relevant block. 
A cell does not belong to the suitable block if and only if its row or its column does not belong to the correct cluster. \Cor{In this section, a toy example where all calculations are possible, is presented. More general simulations are also proposed.}

\Cor{\paragraph{Toy example:}The following simple configurations are considered to illustrate this issue:
\begin{equation}
P_{h,h'}=\begin{pmatrix}\frac{1-x}{2}&\frac{x}{2}\\\frac{x}{2}&\frac{1-x}{2}\\\end{pmatrix}\text{ and }P_{\ell,\ell'}=\begin{pmatrix}\frac{1-y}{2}&\frac{y}{2}\\\frac{y}{2}&\frac{1-y}{2}\\\end{pmatrix}\label{Eq:Model}
\end{equation}
where $x\in[0;1/2]$ and $y\in[0;1/2]$ represent the proportion of misclassified rows and columns respectively and the number of misclassified cells is $x+y-xy$.}

\Cor{The following quantities are equal in the designed model:
\[\mathcal{H}(\boldsymbol{z})=\mathcal{H}(\boldsymbol{z'})=\mathcal{H}(\boldsymbol{w})=\mathcal{H}(\boldsymbol{w'})=2\log2\]
and $\text{coNMI}((\boldsymbol{z},\boldsymbol{w});(\boldsymbol{z}',\boldsymbol{w}'))=\text{ENMI}((\boldsymbol{z},\boldsymbol{w});(\boldsymbol{z}',\boldsymbol{w}'))/2$.\\
}

The ENMI criterion is based on the sum between the NMI criterion on the rows and on the columns. In Figure~\ref{fig:config}, two configurations are shown where the  ENMI criterion takes the value of 1 in the two cases, when the proportion of well classified cells is respectively 50\% (on the left, \Cor{$(x,y)=(0.5,0)$}) and approximately 21\% (on the right, \Cor{$(x,y)\approx(0.1100279,0.1100279)$})\footnote{To find the values in  Figure~\ref{fig:config}, we took Configuration~\eqref{Eq:Model} and looked for the pairs $(x, y)$ so that the ENMI criterion equals $1\pm 10^{- 16}$ by the dichotomy method. As a result, the proportion of misclassified cells and the value of the CARI criterion are slightly approximated.}.  For the left case, the CARI criterion equals $1/3\approx0.33$ and for the right case, it equals approximately  0.53.

Thus, for different values of the rate of misclassified cells, the ENMI value remains unchanged, whereas the CARI criterion is doubtless more sensitive.

\begin{figure}[!h]
    \centering
    \begin{tabular}{c|c}
    \begin{minipage}{0.45\linewidth}
    \[\text{CARI}((z,w),(z',w'))=\frac{1}{3}\approx0.33,\]
    \[1-\text{CE}((z,w),(z',w'))=0.5,\]
    \[\text{NCE}((z,w),(z',w'))=\frac{1}{3}\approx0.33,\]
    \[\text{ENMI}((z,w),(z',w'))=1,\]
    \[\text{coNMI}((z,w),(z',w'))=0.5.\]
    \begin{tikzpicture}
    \draw[blue, ultra thick, dotted] (0,5) -- (2,5);
    \draw[purple, ultra thick] (4,5) -- (2,5);
    \draw[blue, ultra thick, dotted] (0,4.5) -- (1,4.5);
    \draw[purple, ultra thick] (2,4.5) -- (1,4.5);
    \draw[blue, ultra thick, dotted] (2,4.5) -- (3,4.5);
    \draw[purple, ultra thick] (3,4.5) -- (4,4.5);
    
    \draw[green, ultra thick, dotted] (-1,0) -- (-1,2);
    \draw[orange, ultra thick] (-1,4) -- (-1,2);
    \draw[green, ultra thick, dotted] (-0.5,0) -- (-0.5,2);
    \draw[orange, ultra thick] (-0.5,4) -- (-0.5,2);
    
    \draw (4,5) node[right]{$\bw$};
    \draw (4,4.5) node[right]{$\bw'$};
    \draw (-1,0) node[below]{$\bz$};
    \draw (-0.5,0) node[below]{$\bz'$};
    
    \draw[red,opacity=0.3,fill] (1,0) rectangle (3,4);
    
    \draw (0,0) grid [step=0.1] (4,4);
    \draw[thick] (0,0) rectangle (4,4);
    \end{tikzpicture}
    \end{minipage}&
    \begin{minipage}{0.45\linewidth}
    \[\text{CARI}((z,w),(z',w'))\approx0.53,\]
    \[1-\text{CE}((z,w),(z',w'))\approx0.79,\]
    \[\text{NCE}((z,w),(z',w'))\approx0.72,\]
    \[\text{ENMI}((z,w),(z',w'))=1,\]
    \[\text{coNMI}((z,w),(z',w'))=0.5.\]
    \begin{tikzpicture}
    \draw[blue, ultra thick, dotted] (0,5) -- (2,5);
    \draw[purple, ultra thick] (4,5) -- (2,5);
    \draw[blue, ultra thick, dotted] (0,4.5) -- (1.779944,4.5);
    \draw[purple, ultra thick] (2,4.5) -- (1.779944,4.5);
    \draw[blue, ultra thick, dotted] (2,4.5) -- (2.220056,4.5);
    \draw[purple, ultra thick] (2.220056,4.5) -- (4,4.5);
    
    \draw[green, ultra thick, dotted] (-1,0) -- (-1,2);
    \draw[orange, ultra thick] (-1,4) -- (-1,2);
    \draw[green, ultra thick, dotted] (-0.5,0) -- (-0.5,1.779944);
    \draw[orange, ultra thick] (-0.5,2) -- (-0.5,1.779944);
    \draw[green, ultra thick, dotted] (-0.5,2) -- (-0.5,2.220056);
    \draw[orange, ultra thick] (-0.5,4) -- (-0.5,2.220056);
    
    \draw (4,5) node[right]{$\bw$};
    \draw (4,4.5) node[right]{$\bw'$};
    \draw (-1,0) node[below]{$\bz$};
    \draw (-0.5,0) node[below]{$\bz'$};
    
    \draw[red,opacity=0.3,fill] (1.779944,0) rectangle (2.220056,4);
    \draw[red,opacity=0.3,fill] (0,1.779944) rectangle (4,2.220056);
    
    \draw (0,0) grid [step=0.1] (4,4);
    \draw[thick] (0,0) rectangle (4,4);
    \end{tikzpicture}
    \end{minipage}
    \end{tabular}
    \caption{Two configurations with four blocks: partitions in rows and columns are represented by the color lines: for each configuration, one cluster is represented by a solid line and another by a dotted line. The red cells are not in the same blocks for both co-partitions.}
    \label{fig:config}
\end{figure}

\begin{figure}[!h]
\begin{center}
\includegraphics[width=\textwidth]{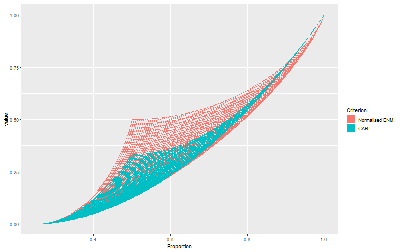}
\caption{\label{Fig:Nappes}\Cor{Evolution of CARI (in purple), $1-\text{CE}$ (in blue), NCE (in green) and ENMI divided by 2, which is equal to coNMI} (in red),  according to the proportion of well classified cells for Configuration~\eqref{Eq:Model} and for $(I,J)=(100000,100000)$.}
\end{center}
\end{figure}

To go further,  Figure~\ref{Fig:Nappes} shows the values of \Cor{ CARI (in purple), $1-\text{CE}$ (in blue), NCE (in green) and ENMI divided by 2, which is equal to coNMI (in red),} according to the proportion of well classified cells for Configuration~\eqref{Eq:Model} and for $I=100000$ and $J=100000$. \Cor{We observe that NCE and CE follow a straight line and are therefore proportional to the number of misclassified cells. Moreover, w}e observe a lower variance for the CARI criterion than for the ENMI criterion; in particular, on the x-axis (for a fixed criterion value, the possible proportions are smaller for the CARI criterion).

\FloatBarrier
\Cor{\paragraph{Simulations:} To compare the criteria in more general cases, we propose the following simulation plan for $(I,J)=(100k,50k)$ with $k\in\{1,2\}$, $H$ and $L$ in the set $\{2,\ldots,8\}$:
\begin{enumerate}
	\item Simulation of a partition $\bz$ with $H$ clusters on $I$ rows: for each row $i\in\{1,\ldots,I\}$ uniform simulation of a value in the set $\{1,\ldots,H\}$. Identical simulation for the partition $\bw$ with $L$ clusters on $J$ columns.
	\item Definition of $\left(\bz^{(0)},\bw^{(0)}\right)=\left(\bz,\bw\right)$.
	\item For $c\in\{1,\ldots,10\,000\}$, do:
	\begin{enumerate}
		\item Simulation of a random variable following a balanced Bernoulli law $Y \sim \mathcal{B}(1/2)$:
		\begin{itemize}
			\item If $Y=0$, let us note $\{u_1,u_2,\ldots,u_{\Gamma}\}$ the ordered set of cluster numbers of the matrix $\bz^{(c-1)} $ and define $\left(\bz^{(c)},\bw^{(c)}\right)=\left(\bz^{(c-1)},\bw^{(c-1)}\right)$. Simulation of an index of the partition $\bz^{(c)}$ following an uniformly variable and simulation of a new value for its cluster in the set $\{u_1,u_2,\ldots,u_{\ell}\}$ following the weights:
			\[\left(\frac{u_{\ell}}{\sum_{\gamma=1}^{\Gamma}u_{\gamma}},\frac{u_{\ell-1}}{\sum_{\gamma=1}^{\Gamma}u_{\gamma}},\ldots,\frac{u_{1}}{\sum_{\gamma=1}^{\Gamma}u_{\gamma}}\right).\]
			\item If $Y=1$, identical procedure except that the drawing is done in the partition of columns $\bw^{(c)}$.
		\end{itemize}
		\item Computation of the values of each criterion between co-partitions $\left(\bz,\bw\right)$ and $\left(\bz^{(c)},\bw^{(c)}\right)$.
		\item If $\text{NCE}\left((\boldsymbol{z},\boldsymbol{w}),\left(\bz^{(c)},\bw^{(c)}\right)\right)<0.01$, break.
	\end{enumerate}
\end{enumerate}
}

\begin{figure}[!ht]
\begin{center}
\begin{tabular}{cccc}
CARI&1-CE&ENMI/2&coNMI\\
\begin{minipage}{0.225\textwidth}
\includegraphics[width=\linewidth]{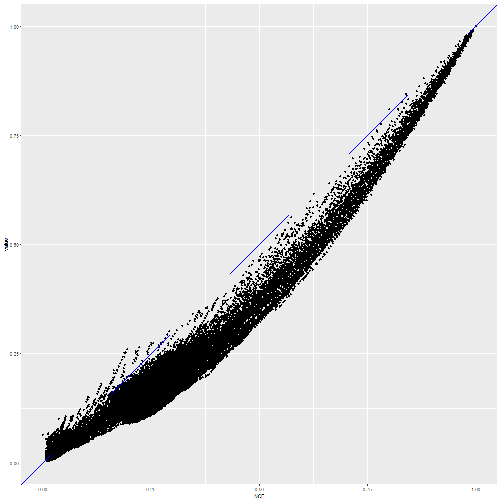}
\end{minipage}&
\begin{minipage}{0.225\textwidth}
\includegraphics[width=\linewidth]{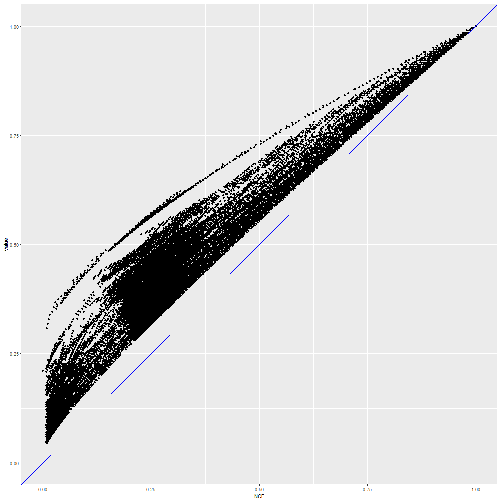}
\end{minipage}&
\begin{minipage}{0.225\textwidth}
\includegraphics[width=\linewidth]{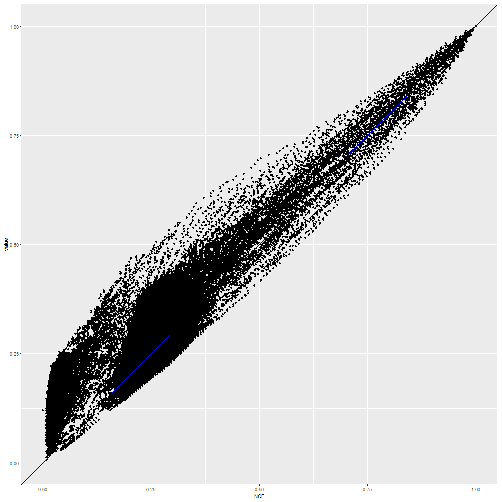}
\end{minipage}&
\begin{minipage}{0.225\textwidth}
\includegraphics[width=\linewidth]{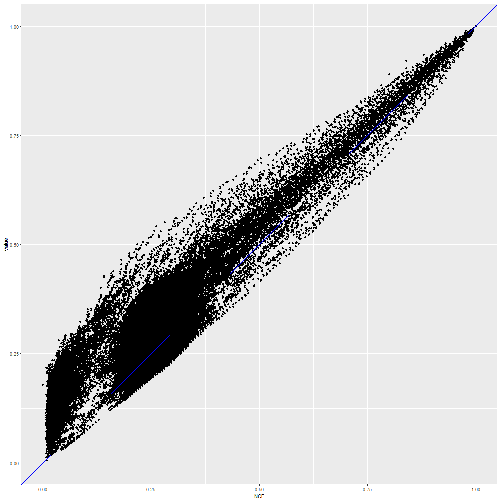}
\end{minipage}\\
\end{tabular}
\caption{\label{fig:comp:simulation}\Cor{Representation of the values of CARI, 1-CE, ENMI/2 and coNMI  for the co-partitions $\left(\bz,\bw\right)$ and $\left(\bz^{(c)},\bw^{(c)}\right)$, according to the value obtained for NCE. The blue line is the bisecting line.}}
\end{center}
\end{figure}

\Cor{Figure~\ref{fig:comp:simulation} shows the values obtained for each criterion as a function of the value obtained by the $NCE$ criterion. }

\Cor{Figure~\ref{fig:comp:simulation} shows that the criterion $CE$ is sensitive to the number of clusters of each partition. The use of the standardized criterion NCE as a reference was therefore preferred. The CARI, ENMI and coNMI criteria do not quite follow the blue line, which means that the values are not perfectly proportional to the number of misclassified cells. Note that the criterion CARI tends to be more penalizing than the other criteria (since its values are below the blue line) but has a lower variance. In contrast, 1-CE, ENMI/2 and coNMI seem to penalize less strongly the proportions of misclassified cells. For the CE criterion, different groups corresponding to different couples $(H, L)$ appear. The comparison of the values obtained can therefore only be made between co-partitions with identical numbers of blocks. Finally, the  ENMI/2 and coNMI criteria are more dispersed than the other criteria and can take a large number of values under the blue line. We observe on the graphs that the same value of the coNMI and ENMI/2 criteria corresponds to a large possibility of proportion of poorly classified cells.}


\subsection{Symmetry}

\Cor{All the criteria are symmetrical.}

\Corr{ The proof of this property regarding the CARI criterion is detailed in Corollary 2.4}.


\Cor{For the CE and NCE criteria, the formula is based on the inverse permutation: as we seek the maximum on all possible permutations, we obtain the result.}

\Cor{Finally, the ENMI and coNMI criteria inherit the symmetry property from the MI criterion.}

\subsection{Boundaries}

\Cor{According to the results presented in Section~\ref{sec:boundaries}, the bound of 1 is reached for two identical co-partitions for the CARI, NCE, coNMI, 1-CE and ENMI/2 criteria.}

\Cor{In contrast, the bound of 0 is reached in the worst cases for the NCE, ENMI and coNMI criteria. In the case of the CARI criterion, the lower bound is slightly negative and tends towards 0 when $I$ and $J$ tend to infinity. On the other hand, the lower bound of the 1-CE criterion is strictly positive as soon as $H$ and $L$ are fixed.}

\subsection{Criteria facing switching label}

\Cor{In classification, clusters are represented by numbers: individuals with the same number are then in the same cluster. The choice of number is generally arbitrary and two partitions can be identical except for a permutation of numbers: this is what we call label switching. It is therefore important that the criterion does not depend on the choice of cluster numbers.}

\Cor{The CARI, ENMI and coNMI criteria are not influenced by switching label as they add up on all the clusters without an imposed order. Changing the order of the clusters does not change the result. In addition, by definition, the CE and NCE criteria seek the optimal value on all permutations. They finally do not to be sensitive to label switching.}

\subsection{Time comparison }
\Cor{ In this part, we compare the execution times of the different implementations on concrete scenarios.}

\begin{figure}[ht!]
\begin{center}
\begin{tabular}{c||c|c}
&$n=250$&$n=500$\\
\hline
\hline
\rotatebox[origin=c]{90}{Professional computer}&\begin{minipage}{0.45\linewidth}
\includegraphics[width=\linewidth]{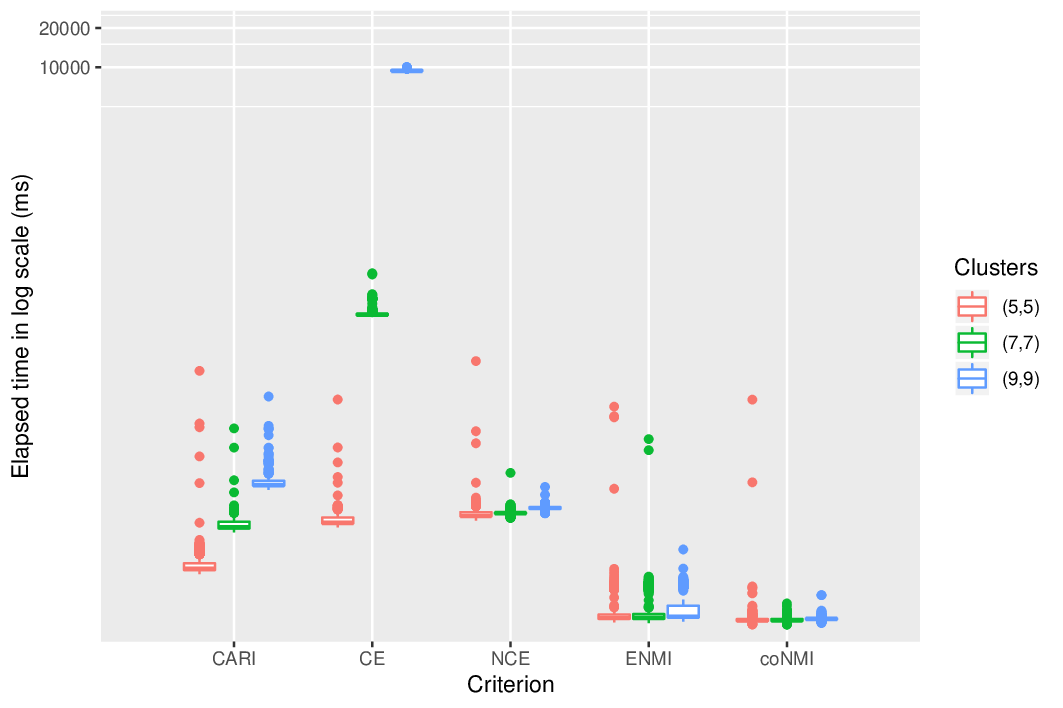}
\end{minipage}&
\begin{minipage}{0.45\linewidth}
\includegraphics[width=\linewidth]{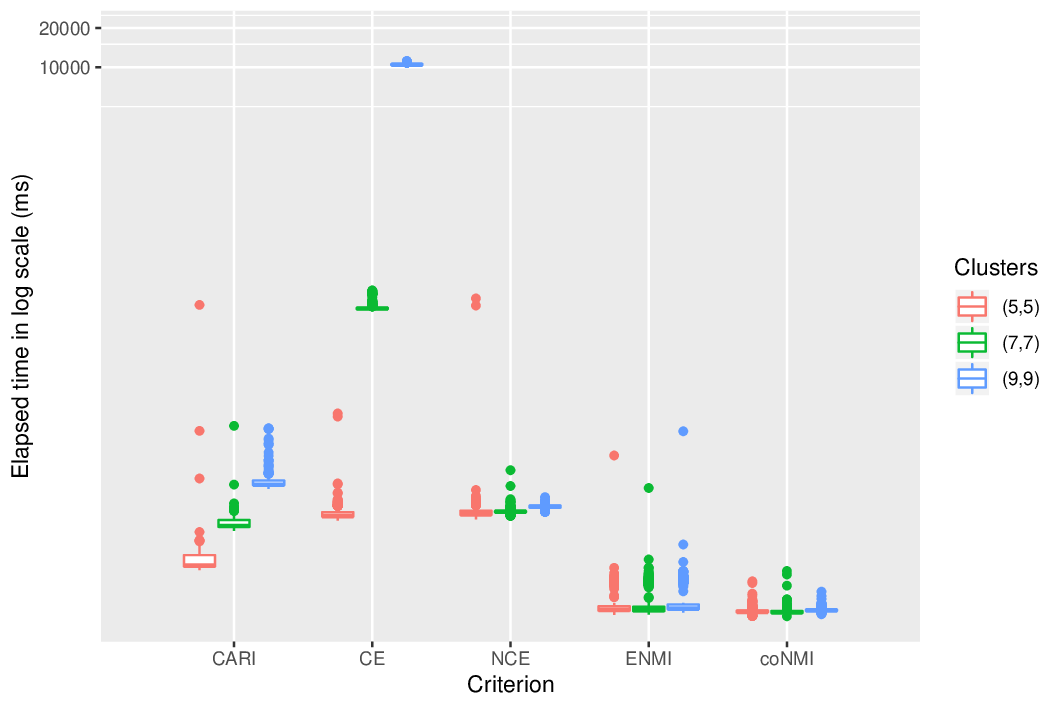}
\end{minipage}\\
\hline
\rotatebox[origin=c]{90}{Server}&\begin{minipage}{0.45\linewidth}
\includegraphics[width=\linewidth]{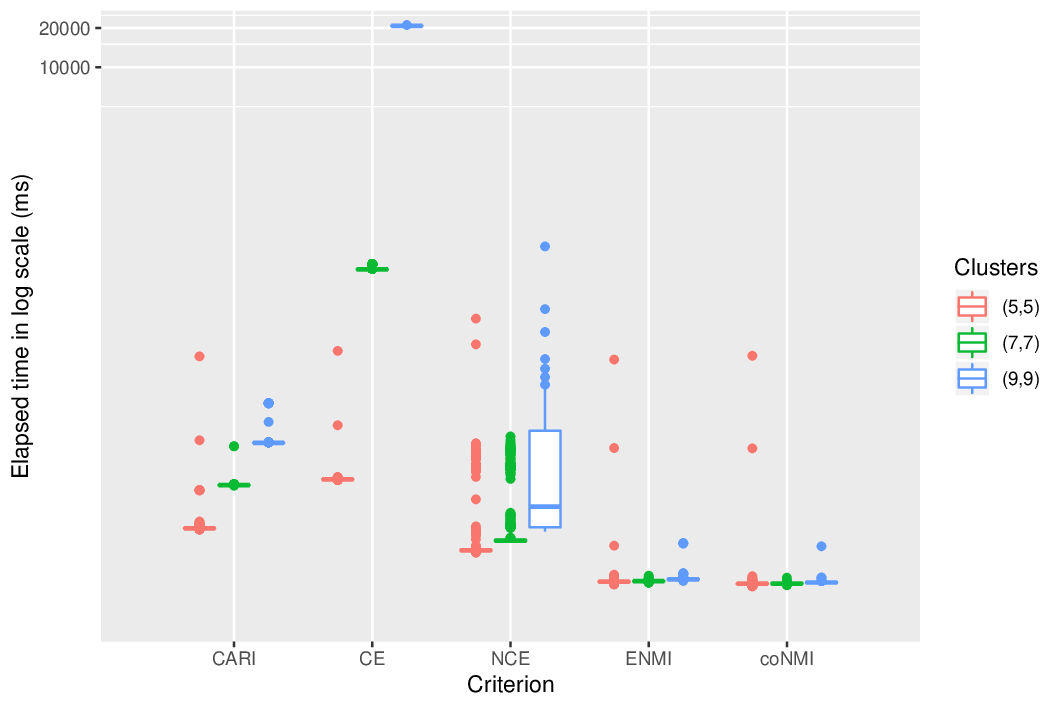}
\end{minipage}&
\begin{minipage}{0.45\linewidth}
\includegraphics[width=\linewidth]{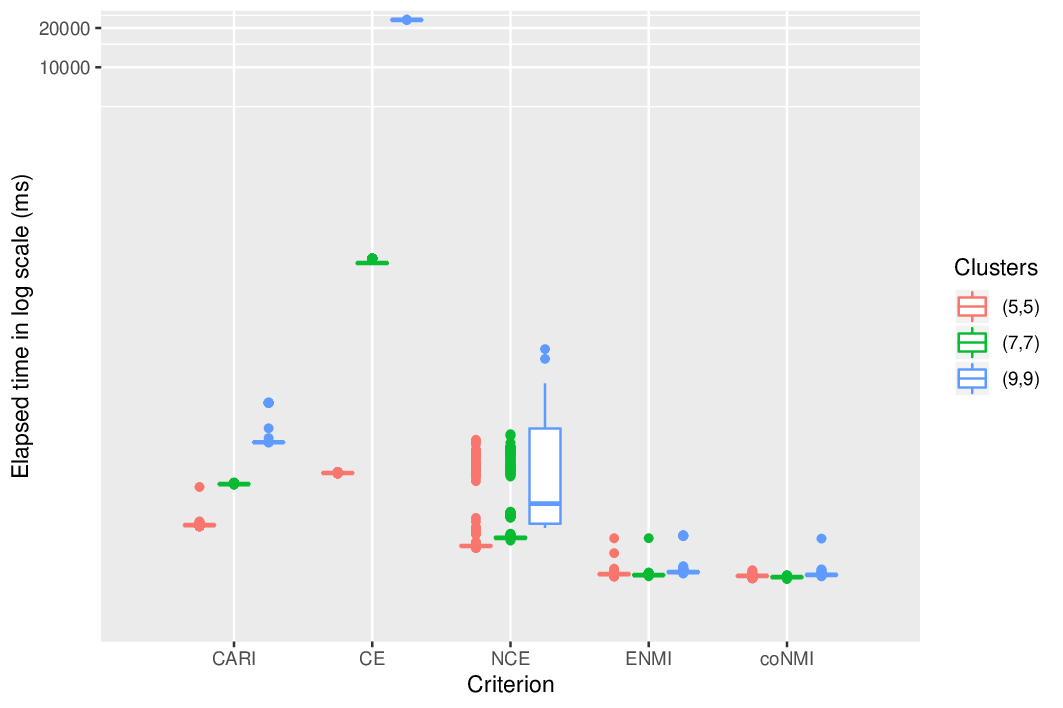}
\end{minipage}\\
\end{tabular}
\end{center}
\caption{Boxplot of the elapsed time computation in log scale in milliseconds, of CARI,  ENMI and CE for different numbers of clusters, $(H,L)\in\{(5,5),(7,7),(9,9)\}$ and $(I,J)=(250,250)$ observations (first column) and $(I,J)=(500,500)$ observations (second column): on a professional computer (windows 10 professional version 1809 workstation with Intel Core i7-8650U CPU 2.11 GHz processor and 32 GB of memory) for the first row and a server (Debian stable amd64 workstation with 2 $\times$ Intel Xeon E5-2640 2.5GHz processor and 64 GB of memory) for the second row.\label{c315}}
\end{figure}

The complexity of the three indices related to the number of observations and the number of clusters is assessed. For this purpose, the procedure is run with $N=1~000$ iterations considering two situations $(I,J)=(250,250)$ observations and 
 $(I,J)=(500,500)$ observations when the number of clusters varies as follows, $(H,L)\in\{(5,5),(7,7),(9,9)\}$. The results obtained with two computer configurations are presented in Figure \ref{c315}. We observe that the elapsed time computation in log scale of the ENMI is the smallest and seems not to be sensitive to the number of clusters or observations. This optimistic result is notably explained by the fact that the ENMI criterion ignores the co-clustering structure in a pair of partitions. The CARI criterion which takes into account this structure, also behaves as desired whatever the number of clusters or observations. For further information, the elapsed time computation of the CARI criterion on a run of the procedure with $N=2000$, $(I,J)=(2000,2000)$ and $(H,L)=(20,20)$, is three seconds on average, which is reasonable in a high-dimensional co-clustering context. On the contrary, the time computation of the CE significantly increases with the number of clusters, which illustrates its dependence on the factorial of this quantity.

\subsection{Comparative study}

\Cor{To conclude this part, we present in Table~\ref{tab:comparaisons} a summary of the previous results.}

\begin{table}[!ht]
\Cor{\caption{Assessment of the behavior of the criteria with regard to the five desired properties. \checkmark means that the criterion is validated while {\xmark} means that the criterion is not fulfilled (or the least efficient compared to the other criteria).\label{tab:comparaisons}}
\begin{tabular}{|p{0.21\textwidth}||*{5}{p{0.13\textwidth}|}}
\cline{2-6}
\multicolumn{1}{c|}{}&CARI&CE&NCE&ENMI&coNMI\\
\multicolumn{1}{c|}{}&&\scriptsize{\citep{Lomet2012selection}}&&\scriptsize{\citep{wyse2016block}}&\\
\hline
1. Proportion of cells misclassified&Moderately&\checkmark&\checkmark&\xmark&\xmark\\
\hline
2. Symmetry&\checkmark&\checkmark&\checkmark&\checkmark&\checkmark\\
\hline
3.1 Maximum limit equal to 1&\checkmark&If we use $1-\text{CE}$&\checkmark&\checkmark&\checkmark\\
\hdashline
3.2 Minimum limit equal to 0&Asympto-tically&\xmark&\checkmark&\checkmark&\checkmark\\
\hline
4. Label switching&\checkmark&\checkmark&\checkmark&\checkmark&\checkmark\\
\hline
5. Execution time&Reasonable&Impossible as soon as \scriptsize{$\max(H,L)>9$}&Reasonable&\checkmark&\checkmark\\
\hline
\end{tabular}}
\end{table}

\Cor{We observe that the CE criterion only checks three properties and cannot be computed in a reasonable execution time (which makes its use often obsolete). The improvement we suggest, namely NCE, is the one that satisfies the most of the properties. The ENMI and coNMI criteria verify the last four properties (in particular they are the fastest in computation time) but offer results less in agreement with the philosophy of co-clustering. Finally, the CARI criterion proposed in this article, does not fully verify all the properties but offers an interesting alternative to the NCE criterion (especially when the number of cells is very large).}

\section{Real data application}
\begin{figure}[ht!]
\begin{center}
\includegraphics[scale=0.7]{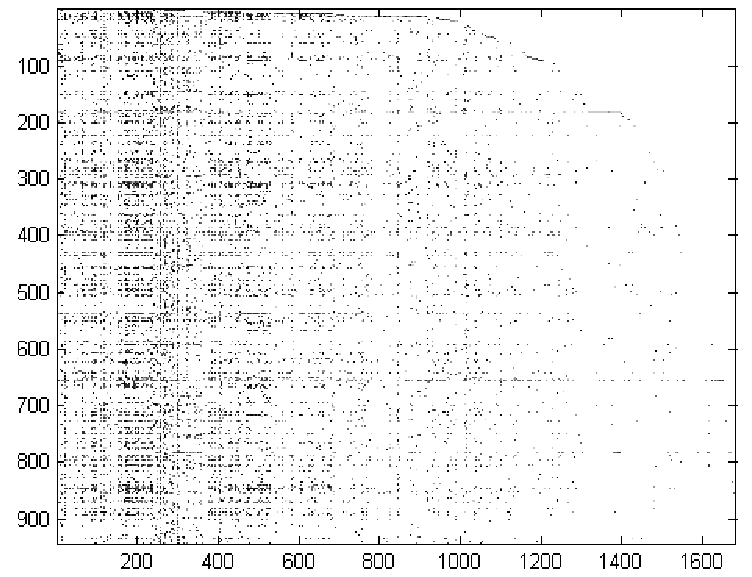}
\caption{ MovieLens data including $100 000$ users ratings from $1$ to $5$, $943$ users and $1682$ films.}\label{movie}
\end{center}
\end{figure}

 In this study, we focus on the MovieLens dataset which includes 100 000 ratings of films from  1 to 5 supplied by 943 users and  involves 1682 films \footnote{\url{http://grouplens.org/datasets/movielens/}}. The studied data matrix (the rows correspond to users and the columns deal with the films) is represented in Figure \ref{movie}. Notice that if data are missing, they are represented by the value 0. In this section, we aim at comparing  co-clustering partitions with high number of clusters, estimated from two methods: the greedy algorithm \textit{GS} proposed by \cite{wyse2016block} and the \textit{Bi-KM1} procedure developed by  \cite{robert2017classification} and implemented in the \url{R}-package \url{bikm1} \citep{robert2020bikm1} available on CRAN. The goal is not to study the methods but to compare the proposed solutions.

%

Thus, the classification goal is to propose groups of individuals sharing the same way to rate specific categories of films. This task is particularly interesting to design recommendation systems.

The co-clustering partitions estimated from the studied real dataset and obtained by the two procedures, are provided by the authors (see Table \ref{movietab}). Both co-clustering partitions have the same order of magnitude regarding the number of row and column clusters.  More precisely, the column clusters could correspond to one dominant category of films such as drama or thriller, but it is not the focus of this section.
As the \textit{Extended Normalized Mutual Information} seems not to be relevant in a co-clustering context, and  the \textit{ classification error} proposed by \cite{Lomet2012selection} cannot be computed in practice, in this case (as the number of row and column clusters is broadly higher than $8$), we easily computed the \textit{Co-clustering Adjusted Rand Index}  between the co-clustering partitions provided by the two procedures. The CARI's value is equal 0.3513 and its weak value shows that the provided solutions are quite far from each other. Like any index, its reach is limited in a real data context, as most of times, the reference partitions are often unknown. But in a context where there are some \textit{a priori} information, thanks to this index, we can easily compare co-clustering methods. Furthermore, this type of index remains essential to assess the performances of a method in a simulation study.
\begin{table}[ht!]
\begin{center}
\begin{tabular}{l|c|c}
& \hspace{0.3cm} $GS$ \hspace{0.3cm} &\textit{Bi-KM1} \\
\hline

 Number of row partitions& 50 & 56 \\

 Number of column partitions&55  &49\\

\end{tabular}

\caption{Characteristics of the co-clustering partitions provided by the two procedures on the MovieLens dataset.}\label{movietab}
\end{center}
\end{table}

%
%
%
%
%
\FloatBarrier
\section{Discussion}
\Cor{In this article, criterion CARI based on the Adjusted Rank Index was introduced. We have shown that it can be computed within a reasonable time, thanks to the use of kronecker products. Its behavior was compared to two criteria already used: the Classification Error used by~\cite{Lomet2012selection} and the Extended Normalized Mutual Information introduced by~\cite{wyse2016block}.\\
We have shown that the classification error criterion varied according to the number of clusters.  The proposed implementation was therefore computationally impossible as soon as the number of clusters exceeds 10. We suggested a new version which allows an estimation independent from the number of clusters. We have proposed a faster implementation by reformulating the issue in terms of an assignment problem.\\
We have shown that, conceptually, the criterion ENMI does not take into account the cell as a statistical individual. A small modification on normalization enables us to propose the coNMI criterion which adresses this weakness.\\
We then compared these criteria regarding five properties that we considered important to measure the agreement between two co-classifications. We have observed that only the NCE criterion offers a value proportional to the number of misclassified cells. The CARI criterion seems to slightly penalize more than the other criteria, but the values seem close to the NCE values. Furthermore, experiments underline the fact that the ENMI and coNM criteria could give the same value for configurations with very different proportions of  misclassified cells. Finally, we have shown that all criteria can be computed within a reasonable time except criterion CE.}

\newpage

\begin{appendix}
\section{Appendix}
\subsection{Proof of Theorem~\ref{th:kro}}
\textbf{Theorem~\ref{th:kro}}. \it We have the following relation,
\begin{eqnarray*}
\boldsymbol{n^{zwz'w'}}=  \boldsymbol{n^{zz'}}\otimes \boldsymbol{n^{ww'}}, \label{kron1}
\end{eqnarray*}

\noindent\textit{where $\otimes$ denotes the Kronecker product between two matrices, $(\boldsymbol{z},\boldsymbol{w})$ and $(\boldsymbol{z'},\boldsymbol{w'})$ are two co-clustering partitions, $\boldsymbol{n^{zwz'w'}}$ is the corresponding contingency table defined according to Definition 2.1, $\boldsymbol{n^{zz'}}$ and $\boldsymbol{n^{ww'}}$ are defined in Section 2.2.}\\


\normalfont
\begin{demo}
Let recall the definition of the Kronecker product.
 Let $\boldsymbol{A}=(a_{i,j})$ be a matrix of size $H\times H'$ and $\boldsymbol{B}$ be a matrix of size $L\times L'$. The Kronecker product is the matrix $\boldsymbol{A\otimes B}$ of size $H\times L$ by $H'\times L'$, defined by successive blocks of size $L\times L'$. The block of the index $i,j$ is equal  $a_{i,j}\times \boldsymbol{B}$:
 \[\boldsymbol{A \otimes B}=\begin{pmatrix}
 a_{1,1} \boldsymbol{B}&\ldots & a_{1,H'} \boldsymbol{B}\\
 \ldots &\ldots & \ldots \\
 a_{H,1} \boldsymbol{B}&\ldots & a_{H,H'} \boldsymbol{B}
 \end{pmatrix}.\]

We started by noting a common trick used in computer science. Indeed, for all $p\in\{1,\ldots,HL\}$, the associated pair $(h,\ell)$ denoting a block of $(\boldsymbol{z,w})$, is respectively the quotient plus 1 and the remainder plus 1 of the Euclidean division of $(p-1)$ by $L$. In other words, we have:
\[(p-1)=(h-1)\times L+(\ell-1).\]

We can easily deduce that there is a bijection between each index  $p$ and the pairs $(h,\ell)$.
In the same way, the assertion is valid for $q$ and the pairs $(h',\ell')$. 

The next lemma is the last step before proving the final result:
\begin{lemma}
For all pairs of indices $p$ and $q$ associated respectively with  blocks $(h,\ell)$ and $(h',\ell')$,
\[n^{zwz'w'}_{p,q}=n_{h,h'}^{zz'}n_{\ell,\ell'}^{ww'}.\]
\end{lemma}
\begin{demo}
We notice that the observation $x_{ij}$ is in block $(h,\ell)$ if and only  if  row $i$ is in  cluster $h$ and column $j$ is in cluster $\ell$.
Thanks to this remark, we can easily see that an observation $x_{ij}$ belongs to block $(h,\ell)$ and  block $(h',\ell')$ if and only if row $i$ belongs at the same time to cluster $h$ and cluster $h'$, and  column $j$  belongs at the same time to cluster $\ell$ and  cluster $\ell'$.

\end{demo}

With the previous results, we finally have:

\begin{eqnarray*}
\boldsymbol{n^{zwz'w'}}&=&\begin{pmatrix}
n_{1,1}^{zwz'w'}&n_{1,2}^{zwz'w'}&\cdots& n_{1,L'}^{zwz'w'}&n_{1,L'+1}^{zwz'w'}&\cdots&n_{1,H'L'}^{zwz'w'}\\
n_{2,1}^{zwz'w'}&n_{2,2}^{zwz'w'}&\cdots& n_{2,L'}^{zwz'w'} &n_{2,L'+1}^{zwz'w'}&\cdots& n_{2,H'L'}^{zwz'w'}\\
\vdots&\vdots&\ddots&&&&\vdots\\
n_{L,1}^{zwz'w'}&n_{L,2}^{zwz'w'}&\cdots&n_{L,L'}^{zwz'w'}&n_{L,L'+1}^{zwz'w'}&\cdots& n_{L,H'L'}^{zwz'w'}\\
n_{L+1,1}^{zwz'w'}&n_{L+1,2}^{zwz'w'}&\cdots& n_{L+1,L'}^{zwz'w'}&n_{L+1,L'+1}^{zwz'w'}&\cdots&n_{L+1,H'L'}^{zwz'w'}\\
\vdots&\vdots&&&&\ddots&\vdots\\
n_{HL,1}^{zwz'w'}&n_{HL,2}^{zwz'w'}&\cdots&n_{HL,L'}^{zwz'w'}&n_{HL,L'+1}^{zwz'w'}&\cdots&n_{HL,H'L'}^{zwz'w'}\\
\end{pmatrix}\\
&=&\begin{pmatrix}
n_{1,1}^{zz'}n_{1,1}^{ww'}&n_{1,1}^{zz'}n_{1,2}^{ww'}&\cdots&n_{1,1}^{zz'}n_{1,H'}^{ww'}&n_{1,2}^{zz'}n_{1,1}^{ww'}&\cdots&n_{1,L'}^{zz'}n_{1,H'}^{ww'}\\
n_{1,1}^{,z'}n_{2,1}^{ww'}&n_{1,1}^{zz'}n_{2,2}^{ww'}&\cdots&n_{1,1}^{zz'}n_{2,H'}^{ww'}&n_{1,2}^{zz'}n_{1,1}^{ww'}&\cdots&n_{1,L'}^{zz'}n_{2,H'}^{ww'}\\
\vdots&\vdots&\ddots&&&&\vdots\\
n_{1,1}^{zz'}n_{H,1}^{ww'}&n_{1,1}^{zz'}n_{H,2}^{ww'}&\cdots&n_{1,1}^{zz'}n_{H,H'}^{ww'}&n_{1,2}^{zz'}n_{1,1}^{ww'}&\cdots&n_{1,L'}^{zz'}n_{H,H'}^{ww'}\\
n_{2,1}^{zz'}n_{1,1}^{ww'}&n_{2,1}^{zz'}n_{1,2}^{ww'}&\cdots&n_{2,1}^{z,z'}n_{1,H'}^{ww'}&n_{2,2}^{zz'}n_{1,1}^{ww'}&\cdots&n_{2,L'}^{zz'}n_{1,H'}^{ww'}\\
\vdots&\vdots&&&&\ddots&\vdots\\
n_{L,1}^{zz'}n_{H,1}^{ww'}&n_{L,1}^{zz'}n_{H,2}^{ww'}&\cdots&n_{L,1}^{zz'}n_{HH'}^{ww'}&n_{L,2}^{zz'}n_{H,1}^{ww'}&\cdots&n_{L,L'}^{zz'}n_{H,H'}^{ww'}\\
\end{pmatrix}\\
&=&\begin{pmatrix}
n_{1,1}^{zz'}\boldsymbol{n^{ww'}}&n_{1,2}^{zz'}\boldsymbol{n^{ww'}}&\cdots&n_{1,L'}^{zz'}\boldsymbol{n^{ww'}}\\
n_{2,1}^{zz'}\boldsymbol{n^{ww'}}&n_{2,2}^{zz'}\boldsymbol{n^{ww'}}&\cdots&n_{2,L'}^{zz'}\boldsymbol{n^{ww'}}\\
\vdots&&\ddots&\vdots\\
n_{L,1}^{zz'}\boldsymbol{n^{ww'}}&n_{L,2}^{zz'}\boldsymbol{n^{ww'}}&\cdots&n_{L,L'}^{zz'}\boldsymbol{n^{ww'}}\\
\end{pmatrix}\\
&=&\boldsymbol{n^{zz'}}\otimes \boldsymbol{n^{ww'}}.\\
\end{eqnarray*}
\end{demo}

\subsection{Proof of Corollary~\ref{cor:kronecker}}

\textbf{Corollary~\ref{cor:kronecker}}. 
\begin{enumerate}
\it
\item $\forall(p,q)\in(H\times L)\times (H'\times L')$, we have the following relations between the margins,

\[n^{zwzw'}_{\centerdot, q}=  n^{zz'}_{\centerdot h'_q}\otimes n^{ww'}_{\centerdot, \ell'_q}
 \text{ and }n^{zwzw'}_{ p, \centerdot }=  n^{zz'}_{h_p, \centerdot }\otimes n^{ww'}_{\ell_p, \centerdot }.\]
\item The CARI associated with the contingency table $\boldsymbol{n^{zwzw'}}$ defined as in Equation (\ref{kron}) remains symmetric, that is to say,
\[\text{CARI}((\boldsymbol{z,w}),(\boldsymbol{z',w'}))=\text{CARI}((\boldsymbol{z',w'}),(\boldsymbol{z,w})).\]
\end{enumerate}
\normalfont
\begin{demo}
\begin{enumerate}
\item This assertion forms part of the known properties of the Kronecker product.
\item The proof of this result is the direct consequence of the following lemma:

\begin{lemma}
We have the following relation, 
\begin{eqnarray*}
\boldsymbol{n^{z'w'zw}}= t(\boldsymbol{n^{zwz'w'}}),\label{kk}
\end{eqnarray*}
where t denotes the transpose of a matrix, $(\boldsymbol{z},\boldsymbol{w})$ and $(\boldsymbol{z'},\boldsymbol{w'})$ are two co-clustering partitions, $\boldsymbol{n^{zwz'w'}}$ is the corresponding contingency table defined according to Theorem 2.3., $\boldsymbol{n^{zz'}}$ and $\boldsymbol{n^{ww'}}$ are defined in Section 2.2.
\end{lemma}
\begin{demo}
Thanks to the property of the Kronecker product with the transpose, we have,
\begin{eqnarray*}
\boldsymbol{n^{z'w'zw}}&=&\boldsymbol{n^{z'z}}\otimes \boldsymbol{n^{w'w}}\\
&=&t(\boldsymbol{n^{zz'}})\otimes t(\boldsymbol{n^{ww'}})\\
&=&t\bigg(\boldsymbol{n^{zz'}}\otimes \boldsymbol{n^{ww'}}\bigg)\\
&=& t(\boldsymbol{n^{zwz'w'}}).
\end{eqnarray*}
\end{demo}

\end{enumerate}
\end{demo}

\subsection{Proof of Proposition~\ref{prop:MI=coMI}}
\textbf{Proposition~\ref{prop:MI=coMI}}. 
\Cor{Given two co-clustering partitions, $(\boldsymbol{z},\boldsymbol{w})$ and $(\boldsymbol{z'},\boldsymbol{w'})$, we have:
\[\text{coMI}((\boldsymbol{z},\boldsymbol{w});(\boldsymbol{z}',\boldsymbol{w}'))=\text{MI}(\boldsymbol{z},\boldsymbol{z'})+\text{MI}(\boldsymbol{w},\boldsymbol{w'}).\]}

\begin{demo}
\Cor{Given two co-clustering partitions, $(\boldsymbol{z},\boldsymbol{w})$ and $(\boldsymbol{z'},\boldsymbol{w'})$, and thanks to the results of Theorem~\ref{th:kro}, we observe that for all $(p,q)\in\{1,\ldots,HL\}\times \{1,\ldots,H'L'\}$, we have:
\begin{eqnarray*}
\frac{n_{p,q}^{zwz'w'}}{IJ}&=&\frac{n_{h_p,h'_q}^{zz'}n_{\ell_p,\ell_q}^{ww'}}{IJ}\text{ by Theorem~\ref{th:kro},}\\
&=&P_{h_p,h'_q}P_{\ell_p,\ell'_q}\text{ defined in Section~\ref{sec:critENM},}
\end{eqnarray*}}
\Cor{where $(h_p,\ell_p)$ (resp. $(h'_q,\ell'_q)$) are the coordinates in the co-clustering partition $(\bz,\bw)$ (resp. $(\bz',\bw')$) associated with the block $p$ (resp. $q$). For the same reason and thanks to the results of Corollary~\ref{cor:kronecker}, we have:
\[\frac{n^{zwz'w'}_{ p, \centerdot }}{IJ}=P_{h_p}P_{\ell_p}\text{ and }\frac{n^{zwz'w'}_{ \centerdot,q }}{IJ}=P_{h'_q}P_{\ell'_q}.\]}
\Cor{With these findings, we have:
\begin{eqnarray*}
&&\!\!\!\!\!\!\!\!\!\!\text{coMI}((\boldsymbol{z},\boldsymbol{w});(\boldsymbol{z}',\boldsymbol{w}'))\\
&=&\sum_{p,q}\frac{n_{p,q}^{zwz'w'}}{IJ}\log\left(\frac{n_{p,q}^{zwz'w'}IJ}{n^{zwz'w'}_{ p, \centerdot }n^{zwz'w'}_{ \centerdot,q }}\right)\\
&=&\sum_{p,q}\frac{n_{p,q}^{zwz'w'}}{IJ}\log\left(\frac{n_{p,q}^{zwz'w'}}{IJ}\frac{IJ}{n^{zwz'w'}_{ p, \centerdot }}\frac{IJ}{n^{zwz'w'}_{ \centerdot,q }}\right)\\
&=&\sum_{h_p,h'_q,\ell_p,\ell'_q}P_{h_p,h'_q}P_{\ell_p,\ell'_q}\log\left(P_{h_p,h'_q}P_{\ell_p,\ell'_q}\frac{1}{P_{h_p}P_{\ell_p}}\frac{1}{P_{h'_q}P_{\ell'_q}}\right)\\
&=&\sum_{h_p,h'_q,\ell_p,\ell'_q}P_{h_p,h'_q}P_{\ell_p,\ell'_q}\left[\log\left(\frac{P_{h_p,h'_q}}{P_{h_p}P_{h'_q}}\right)+\log\left(\frac{P_{\ell_p,\ell'_q}}{P_{\ell_p}P_{\ell'_q}}\right)\right]\\
&=&\sum_{h_p,h'_q}P_{h_p,h'_q}\log\left(\frac{P_{h_p,h'_q}}{P_{h_p}P_{h'_q}}\right)\underbrace{\sum_{\ell_p,\ell'_q}P_{\ell_p,\ell'_q}}_{=1}+\sum_{\ell_p,\ell'_q}P_{\ell_p,\ell'_q}\log\left(\frac{P_{\ell_p,\ell'_q}}{P_{\ell_p}P_{\ell'_q}}\right)\underbrace{\sum_{h_p,h'_q}P_{h_p,h'_q}}_{=1}\\
&=&\text{MI}(\boldsymbol{z},\boldsymbol{z'})+\text{MI}(\boldsymbol{w},\boldsymbol{w'}).\\
\end{eqnarray*}}
\end{demo}

\subsection{Proof of the complexities}\label{section:Appendix:Complexity}

\paragraph{Complexity of ENMI and coNMI.} For the complexity of ENMI, we explain the computational cost for $NMI(\bz,\bz')$ as follows:
\begin{enumerate}
    \item The computation of $(P_{h,h'})_{1\leq h\leq H, 1\leq h'\leq H'}$: $\mathcal{O}\left(I\right)$ because we look for each row $i$ of the initial matrix to which cell $(h,h')$ of the contingency table it belongs to and we add 1 to its value.
    \item For the same reasons, the computation of $(P_{h})_{1\leq h\leq H}$ (resp. $(P_{h'})_{1\leq h'\leq H'}$) is $\mathcal{O}\left(HI\right)$.
    \item Finally, the computation of the criterion $NMI(\bz,\bz')$ given the previous matrices is a sum on each pair $(h,h')\in\{1,\ldots,H\}\times\{1,\ldots,H'\}$ so this step has a complexity of $\mathcal{O}\left(HH'\right)$.
\end{enumerate}

So, the complexity of the $NMI(\bz,\bz')$ computation is $\mathcal{O}\left(max(HH',I)\right)$.

As there is a similar complexity for the computation of $NMI(\bw,\bw')$, we have the result.

\paragraph{Complexity of CARI.} For the complexity of CARI, we proceed in three steps:
\begin{enumerate}
    \item As $\boldsymbol{n^{zz'}}=(IP_{h,h'})_{1\leq h\leq H, 1\leq h'\leq H'}$, the complexity is $\mathcal{O}\left(I\right)$. It is the same reasoning for the computation of $\boldsymbol{n^{zwz'w'}}$.
    \item Thanks to the equation~\ref{kron}, the computation of $\boldsymbol{n^{zwz'w'}}$ is only the Kronecker product of the matrices $\boldsymbol{n^{zz'}}$ and $\boldsymbol{n^{ww'}}$. The complexity is $\mathcal{O}\left(HH'LL'\right)$.
    \item Given the matrices $\boldsymbol{n^{zz'}}$, $\boldsymbol{n^{ww'}}$ and $\boldsymbol{n^{zwz'w'}}$, the most time-consuming operation of the CARI's computation is $\sum_{p,q}\binom{n_{p,q}^{zwz'w'}}{2}$ what has a complexity of $\mathcal{O}\left(HH'LL'\right)$.
\end{enumerate}
Finally, the complexity is $\mathcal{O}\left(\max\left(HH'LL',I,J\right)\right)$.

\paragraph{Complexity of CE.} For the complexity of $CE$, we have:
\begin{enumerate}
    \item For each permutation $\sigma\in \mathfrak{S}(\{1,...,H\})$, we compute:
    \begin{itemize}
        \item $\frac{1}{I}\sum_{i,h}z_{ih}z'_{i\sigma(h)}$ with a complexity of $\mathcal{O}\left(HI\right)$.
    \end{itemize}
    As the cardinal number of $\mathfrak{S}(\{1,...,H\})$ is $H!$, we obtain a complexity of $\mathcal{O}\left(H!HI\right)$.
    \item We have a similar result for the computation of $\text{dist}_{\scriptsize{J,L}}(\boldsymbol{w},\boldsymbol{w'})$; that is to say, we have $\mathcal{O}\left(L!LJ\right)$.
    \item Given $\text{dist}_{\scriptsize{I,H}}(\boldsymbol{z},\boldsymbol{z'})$ and $\text{dist}_{\scriptsize{J,L}}(\boldsymbol{w},\boldsymbol{w'})$, the computation of $CE$ is only three operations.
\end{enumerate}

At the end, the complexity of $CE$ is $\mathcal{O}\left(max\left(H!HI,L!LJ\right)\right)$.

\Cor{\paragraph{Complexity of NCE.} For the complexity of $NCE$, we use the hungarian algorithm with a quartic complexity:
\begin{enumerate}
    \item The complexity to cumpute the contingency table $\boldsymbol{n^{zz'}}$ is of the order of $\mathcal{O}\left(I\right)$ as stated previously.
    \item optimizing the strongest diagonal using the Hungarian algorithm has a complexity of $\mathcal{O}\left(\max(H^4,H'^4)\right)$.
\end{enumerate}
Thus, computing the distance of the partitions $\bz$ and $\bz'$,has a complexity of $\mathcal{O}\left(\max(H^4,H'^4,I)\right)$. We have the similar result for the partition on  columns and the result of the global complexity is proved.}

\subsection{Proof of the boundaries}
\label{section:Appendix:Boundarie}
\paragraph{Boundaries of ENMI.} For the boundaries of ENMI, we can observe that $MI(\bz,\bz')$ is the Kullback-Leibler divergence between the empirical distribution of the joint distribution and the product of the empirical marginal distributions. 
\subparagraph{Lower bound.} In particular, we know that the Kullback-Leibler divergence is greater than $0$; obtained in the case where the empirical density $(P_{h,h'})_{1\leq h\leq H, 1\leq h'\leq H'}$ is the product of the two empirical marginal distributions:
\[\forall (h,h')\in\{1,\ldots,H\}\times\{1,\ldots,H'\}, P_{h,h'}=P_{h}P_{h'}.\]
In this case, we obtain:
\begin{eqnarray*}
\text{MI}(\bz,\bz')&=&\sum_{h,h'}P_{h,h'}\log\left(\frac{P_{h,h'}}{P_{h}P_{h'}}\right)\\
&=&\sum_{h,h'}P_{h,h'}\log\underbrace{\left(\frac{P_{h}P_{h'}}{P_{h}P_{h'}}\right)}_{=1}\\
&=&0.
\end{eqnarray*}
The lower bound of $\text{EMMI}((\bz,\bw),(\bz',\bw'))$ is a consequence of the sum of $\text{MI}(\bz,\bz')$ and $\text{MI}(\bw,\bw')$ whose are both positive.
\subparagraph{Upper bound on MI.} At the opposite, the upper bound is obtained when $\bz$ is a sub-partition of $\bz'$ if $H>H'$ (or inversely). In this case, for each $h\in\{1,\ldots,H\}$, there exists $h'\in\{1,\ldots,H'\}$ so that for all $i\in\{1,\ldots,I\}$, $z_{ih}=1$ implies that $z_{ih'}=1$; in this case, we say that $h$ is in the subset $S_{h'}$ and we have:
\begin{eqnarray*}
P_{h,h'}&=&\frac{1}{I}\sum_{i=1}^I\mathds{1}_{\{z_i=h,z'_{i}=h'\}}\\
&=&P_h\mathds{1}_{\{h\in S_{h'}\}}
\end{eqnarray*}
and we have:
\begin{eqnarray*}
\sum_{h=1}^HP_{h,h'}&=&\sum_{h=1}^HP_h\mathds{1}_{\{h\in S_{h'}\}}\\
&=&\sum_{h=1}^H\left(\frac{1}{I}\sum_{i=1}^I\mathds{1}_{\{z_i=h\}}\right)\mathds{1}_{\{h\in S_{h'}\}}\\
&=&\frac{1}{I}\sum_{i=1}^I\sum_{h=1}^H\mathds{1}_{\{z_i=h\}}\mathds{1}_{\{h\in S_{h'}\}}\\
&=&\frac{1}{I}\sum_{i=1}^I\mathds{1}_{\{z_i=h'\}}\\
&=&P_{h'}.
\end{eqnarray*}
Then, we have:
\begin{eqnarray*}
MI(\bz,\bz')&=&\sum_{h'=1}^{H'}\sum_{h=1}^HP_{h,h'}\log\left(\frac{P_{h,h'}}{P_{h}P_{h'}}\right)\\
&=&\sum_{h'=1}^{H'}\sum_{h=1}^HP_h\mathds{1}_{\{h\in S_{h'}\}}\log\left(\frac{P_{h}}{P_{h}P_{h'}}\right)\\
&=&\sum_{h'=1}^{H'}\log\left(\frac{1}{P_{h'}}\right)\sum_{h=1}^HP_h\mathds{1}_{\{h\in S_{h'}\}}\\
&=&-\sum_{h'=1}^{H'}\left(\log P_{h'}\right)P_{h'}\\
&=&\mathcal{H}(\bz')\\
&=&\min\left(\mathcal{H}(\bz),\mathcal{H}(\bz')\right).\\
\end{eqnarray*}

The last equation is a consequence of the fact that $\bz$ is a sub-partition of $\bz'$.

As the minimum is less than the maximum, we observe that $\text{MI}(\bz,\bz')$ is smaller than $1$, the $\text{EMMI}((\bz,\bw),(\bz',\bw'))$ is then smaller than $2$.

\paragraph{Boundaries of CARI.} To prove the upper bound of the lower bound, we use the case where the partitions $\bz$, $\bz'$, $\bw$ and $\bw'$ are equidistributed and the intersections of a cluster of $\bz$ (resp. $\bw$) with each cluster of $\bz'$ (resp. $\bw'$) have the same number of rows (resp. columns). In particular, we have:
\[n_{p,\cdot}^{\bz\bw\bz'\bw'}=\frac{IJ}{HL},\,n_{\cdot,q}^{\bz\bw\bz'\bw'}=\frac{IJ}{H'L'}\text{ and }n_{p,q}^{\bz\bw\bz'\bw'}=\frac{IJ}{HH'LL'}.\]
With this configuration, we have:
\begin{eqnarray*}
&&\!\!\!\!\!CARI((\bz,\bz'),(\bw,\bw'))\\
&=&\frac{\sum_{p,q}\frac{\frac{IJ}{HH'LL'}\left(\frac{IJ}{HH'LL'}-1\right)}{2}-\left(\sum_{p}\frac{\frac{IJ}{HL}\left(\frac{IJ}{HL}-1\right)}{2}\right)\left(\sum_{q}\frac{\frac{IJ}{H'L'}\left(\frac{IJ}{H'L'}-1\right)}{2}\right)\times\frac{2}{IJ(IJ-1)}}{\frac{1}{2}\left[\sum_{p}\frac{\frac{IJ}{HL}\left(\frac{IJ}{HL}-1\right)}{2}+\sum_{q}\frac{\frac{IJ}{H'L'}\left(\frac{IJ}{H'L'}-1\right)}{2}\right]-\left(\sum_{p}\frac{\frac{IJ}{HL}\left(\frac{IJ}{HL}-1\right)}{2}\right)\left(\sum_{q}\frac{\frac{IJ}{H'L'}\left(\frac{IJ}{H'L'}-1\right)}{2}\right)\times\frac{2}{IJ(IJ-1)}}\\
&=&\frac{IJ\left(\frac{IJ}{HH'LL'}-1\right)-\left(\frac{IJ}{HL}-1\right)\left(IJ\left(\frac{IJ}{H'L'}-1\right)\right)\times\frac{1}{IJ-1}}{\frac{1}{2}\left[IJ\left(\frac{IJ}{HL}-1\right)+IJ\left(\frac{IJ}{H'L'}-1\right)\right]-\left(\frac{IJ}{HL}-1\right)\left(IJ\left(\frac{IJ}{H'L'}-1\right)\right)\times\frac{1}{IJ-1}}\\
&=&\frac{\frac{IJ}{HH'LL'}-1-\left(\frac{IJ}{HL}-1\right)\left(\frac{IJ}{H'L'}-1\right)\times\frac{1}{IJ-1}}{\frac{IJ}{2HL}+\frac{IJ}{2H'L'}-1-\left(\frac{IJ}{HL}-1\right)\left(\frac{IJ}{H'L'}-1\right)\times\frac{1}{IJ-1}}.\\
\end{eqnarray*}

For the conjecture, we have:
\begin{eqnarray*}
&&\!\!\!\!\!\!\frac{\frac{IJ}{HH'LL'}-1-\left(\frac{IJ}{HL}-1\right)\left(\frac{IJ}{H'L'}-1\right)\times\frac{1}{IJ-1}}{\frac{IJ}{2HL}+\frac{IJ}{2H'L'}-1-\left(\frac{IJ}{HL}-1\right)\left(\frac{IJ}{H'L'}-1\right)\times\frac{1}{IJ-1}}\\
&=&\frac{\frac{IJ}{HH'LL'}-1-\left(\frac{IJ}{HL}-1\right)\left(\frac{IJ}{H'L'}-1\right)\left[\frac{1}{IJ}+o\left(\frac{1}{IJ}\right)\right]}{\frac{IJ}{2HL}+\frac{IJ}{H'L'}-1-\left(\frac{IJ}{HL}-1\right)\left(\frac{IJ}{2H'L'}-1\right)\left[\frac{1}{IJ}+o\left(\frac{1}{IJ}\right)\right]}\\
&=&\frac{\frac{IJ}{HH'LL'}-1-\frac{IJ}{HLHL'}+\frac{1}{HL}+\frac{1}{H'L'}+o(1)}{\frac{IJ}{2HL}+\frac{IJ}{H'L'}-1-\frac{IJ}{HH'LL'}+\frac{1}{HL}+\frac{1}{H'L'}+o\left(1\right)}\\
&=&\frac{-1+\frac{1}{HL}+\frac{1}{H'L'}}{\frac{IJ}{HLH'L'}\left(\frac{HL+H'L'}{2}-1\right)-1+\frac{1}{HL}+\frac{1}{H'L'}}+o(1)\\
&=&\frac{HL+H'L'-HLH'L'}{IJ\left(\frac{HL+H'L'}{2}-1\right)+HL+H'L'-HLH'L'}+o(1).\\
\end{eqnarray*}

\subparagraph{Upper bound.} As for every partition $\bz$, $\bz'$, $\bw$ and $\bw'$, we have: \[\sum_{p,q}\binom{n_{p,q}^{zwzw'}}{2}\leq \frac{1}{2}\left[\sum_{p}\binom{n_{p,\cdot}^{zwz'w'}}{2}+\sum_{q}\binom{n_{\cdot,q}^{zwz'w'}}{2}\right]\]
then the upper bound is $1$.

\paragraph{Boundaries of (N)CE.} \Cor{Let's start by noting that $\sum\limits_{i,h}z_{ih}z'_{i\sigma(h)}$ is smaller than $I$ for each permutation $\sigma\in \mathfrak{S}(\{1,...,H\})$. In this demonstration, we assume that $H\geq H'$ and if $H'<H$ then we complete partition $\boldsymbol{z'}$ with empty clusters such that the matrix~$\boldsymbol{n}^{\boldsymbol{zz'}}$ is square. Thanks to the formulation under an assignment problem, we have shown that minimizing the criterion in Equation~\eqref{Eq:dist_CE} leads to seek the optimal permutation on the columns of the matrix~$\boldsymbol{n}^{\boldsymbol{zz'}}$ which would provide the diagonal with the highest sum. Given a permutation $\sigma\in \mathfrak{S}(\{1,...,H\})$, $\boldsymbol{n}^{\sigma}$ denotes the matrix~$\boldsymbol{n}^{\sigma}=\left(\boldsymbol{n}_{h\sigma(h')}^{\boldsymbol{zz'}}\right)_{1\leq h,h'\leq H}$ after permutation and $S_{\sigma}=\sum\limits_{h=1}^H\boldsymbol{n}_{hh}^{\sigma}$ the sum of the cells on the diagonal $D_{\sigma}=\left(\boldsymbol{n}^{\sigma}_{11},\ldots,\boldsymbol{n}^{\sigma}_{HH}\right)$. Without loss of generality, we assume that the optimal permutation $\sigma$ is identity $\Id$ and we search the worst possible value for $S_{\Id}$ which means that:
\begin{itemize}
	\item The value of $S_{\Id}$ must be as small as possible.
	\item For each permutation $\sigma\in \mathfrak{S}(\{1,...,H\})$, $S_{\Id}\geq S_{\sigma}$.
\end{itemize}
If $\sigma_h$ denotes the circular permutation such that for all $h'\in \{1,\ldots,H\}$,
\[\sigma_h(h')\equiv h'+h \pmod H,\]
the n-tuple $\left(D_{\Id},D_{\sigma_1},\ldots,D_{\sigma_{H-1}}\right)$ is a partition of the cells of matrix~$\boldsymbol{n}^{\boldsymbol{zz'}}$. It means that,
\[S_{\Id}+\sum_{h=1}^{H-1}S_{\sigma_h}=I.\]
As a result, we have:
\begin{eqnarray*}
I&=&S_{\Id}+\sum_{h=1}^{H-1}S_{\sigma_h}\\
&\leq&S_{\Id}+\sum_{h=1}^{H-1}S_{\Id}\\
&\leq&HS_{\Id},\\
\end{eqnarray*}
that is to say $S_{\Id}\geq I/H$ and the equality is reached if all the cells of matrix~$\boldsymbol{n}^{\boldsymbol{zz'}}$ are identical.}

\Cor{In the end, for any permutation $\sigma\in \mathfrak{S}(\{1,...,H\})$, we have:
\begin{eqnarray*}
\frac{I}{H}\leq \sum_{h}\boldsymbol{n}_{h\sigma(h)}^{\boldsymbol{zz'}}\leq I&\Leftrightarrow&\frac{I}{H}\leq \sum_{i,h}z_{ih}z'_{i\sigma(h)}\leq I\\
&\Leftrightarrow&\frac{1}{H}\leq \frac{1}{I}\sum_{i,h}z_{ih}z'_{i\sigma(h)}\leq 1\\
&\Leftrightarrow&\frac{1}{H}\leq 
\max_{\sigma\in \mathfrak{S}(\{1,...,H\})}\frac{1}{I}\sum_{i,h}z_{ih}z'_{i\sigma(h)}\leq 1\\
&\Leftrightarrow&1-\frac{1}{H}\geq 
\max_{\sigma\in \mathfrak{S}(\{1,...,H\})}\frac{1}{I}\sum_{i,h}z_{ih}z'_{i\sigma(h)}\geq 0.\\
\end{eqnarray*}
The last inequality enables us to conclude.}

\Cor{In the case where $I$ is not divisible by $H^2$, the bound can be improved by regularly distributing the numbers.}

\end{appendix}

\bibliographystyle{newapa}

\bibliography{Biblio}

\end{document}